\documentstyle[12pt,epsfig]{article}
\begin{document}

\begin{titlepage}
\rightline{February 2015}
\vskip 2cm
\centerline{\Large \bf
Dissipative dark matter explains rotation curves}

\vskip 2.2cm
\centerline{R. Foot\footnote{
E-mail address: rfoot@unimelb.edu.au}}

\vskip 0.7cm
\centerline{\it ARC Centre of Excellence for Particle Physics at the Terascale,}
\centerline{\it School of Physics, University of Melbourne,}
\centerline{\it Victoria 3010 Australia}
\vskip 2cm
\noindent
Dissipative dark matter, where dark matter particles interact with a massless (or very light)
boson, is studied. Such dark matter can arise in simple hidden sector gauge models, including
those featuring an unbroken $U(1)'$ gauge symmetry, leading to a dark photon.
Previous work has shown that such models can not only explain the LSS and CMB, but 
potentially also dark matter phenomena on small scales, such as the inferred cored structure of dark matter halos. 
In this picture, dark matter halos of disk galaxies not only cool via dissipative interactions
but are also heated via ordinary supernovae
(facilitated by an assumed photon - dark photon kinetic mixing interaction).
This interaction between the dark matter halo and ordinary baryons, a very special feature of these
types of models, plays a critical role in governing the physical properties of the dark
matter halo. 
Here, we further study the implications of this type of dissipative dark matter
for disk galaxies. 
Building on earlier work, we develop a simple formalism which aims to describe the effects of 
dissipative dark matter in a fairly model independent way. This formalism is then applied to generic 
disk galaxies.
We also consider specific examples, including NGC 1560 and a sample 
of dwarf galaxies from the LITTLE THINGS survey. 
We find that dissipative dark matter, as developed here, does a fairly good job accounting for 
the rotation curves of the galaxies considered. Not only does dissipative dark matter explain the
linear rise of the rotational velocity of dwarf galaxies at small radii, but it can   
also explain the observed wiggles in rotation curves which are 
known to be correlated with corresponding features
in the disk gas distribution.

\end{titlepage}

\section{Introduction}

\vskip 0.3cm

The evidence for the existence of nonbaryonic dark matter in the Universe has grown steadily over the
years. In particular, dark matter is needed to explain the observed large scale structure and 
Cosmic Microwave Background radiation (CMB). The angular power spectrum of the anisotropy of the CMB  
has been measured with impressive precision by the Planck satellite \cite{planck}, South Pole telescope \cite{SPT}, 
WMAP \cite{WMAP} and many other instruments. Cold dark matter models, both collisionless \cite{dodelson}
and even those with nontrivial self-interactions, e.g. \cite{berezhiani,foot13,canada},
provide a simple explanation for this data. 
On much smaller scales, dark matter is also implied by measurements of 
the rotation curves of disk galaxies \cite{rubin}. Typically such curves are asymptotically flat out to the visible
edge of galaxies, a behavior in sharp contrast to the Keplerian decline anticipated from Newton's 
law of gravity.
 
Although
collisionless cold dark matter does an excellent job in explaining the large scale structure and CMB,
it is much less successful in explaining the measured galactic rotation curves, e.g. \cite{rot0,rot1,rot2,rot3}.
In particular, rotation curve data prefer dark matter  
with a cored distribution rather than the cuspy profile predicted from collisionless cold dark matter simulations \cite{nfw}
(for a review and more detailed bibliography see also \cite{review}).
However, dark matter with nontrivial particle physics properties can also explain large scale structure and CMB,
and these nontrivial properties might play
important roles on galactic and subgalactic scales.
If this is the case, then rotation curves might contain important clues 
(such as the aforementioned preference for cored dark matter profiles)
which can help unravel the mystery of dark matter. 

It has been argued in previous work \cite{footvolkas,foot1,foot2,foot3,footsil,foot4} that certain dissipative dark matter
models can in fact lead to dark matter halos consistent with observations.
The general idea is that the dark matter halo of a disk galaxy
takes the form of a dark plasma, with dark matter particles
self interacting via a massless dark photon. 
Such a scenario can naturally arise if dark matter results from a hidden sector 
of particles and forces,
where the hidden sector contains an unbroken $U(1)'$ (dark electromagnetism) 
gauge interaction.
In such a picture the halo can be modeled as a fluid, governed
by Euler's equations. The halo has substantial cooling from processes such as dark bremsstrahlung.
Heating is also possible. Within this framework 
it was argued that ordinary core collapse supernovae could
supply a substantial halo heat source. This mechanism requires only a  tiny interaction
coupling ordinary particles (electrons and positrons produced in the supernova core) 
to light MeV scale dark matter particles, with 
kinetic mixing of the photon
with the dark photon identified as a theoretically favored candidate.
This energy is
further processed into dark photons
in the region around the supernova; the end result is that a 
significant fraction of the supernova's core collapse energy 
can be transmitted to the halo via these dark photons.

In this scenario the dark matter halo is a complicated dynamical system.
Considerable simplification is possible if the system evolves to a
steady-state configuration where
heating and cooling rates approximately balance.
It was argued in previous work that such
a picture can explain why dark matter forms a cored profile in disk galaxies.
Schematically, the cooling rate from dark bremsstrahlung and other dissipative
interactions at a particular location, $P$, is proportional to the square of the dark matter number density, 
$n({\bf r})$,
while the heating rate is proportional to the product of the number density and 
dark photon energy flux at that location, $F_{\gamma_D}({\bf r})$. The particular process considered being dark photoionization
with cross-section, $\sigma_{DP}$.  That is,
\begin{eqnarray}
d\Gamma_{cool} &=& \Lambda n({\bf r})^2 \ dV \ ,
\nonumber \\
d\Gamma_{heat} &=& n({\bf r})  F_{\gamma_D}({\bf r}) \sigma_{DP} \ dV
\label{1z}
\end{eqnarray}
where $\Lambda$ is a constant which depends on the parameters of the particular dissipative model.
Given that the dark photon flux produced from ordinary supernovae will have some frequency
spectrum, the above cross section represents an appropriately  weighted frequency average.

Matching the heating and cooling rates of Eq.(\ref{1z}) implies that
\begin{eqnarray}
n({\bf r}) = {F_{\gamma_D}({\bf r}) \ \sigma_{DP} \over \Lambda}
\label{2z}
\ .
\end{eqnarray}
At distances $r \gg r_D$ (where $r_D$ is the exponential disk scale length associated with the stars: 
$\Sigma^* = {m_D \over 2\pi r_D^2} e^{-r/r_D}$)
the flux of dark photons falls at the `geometric' rate
$F_{\gamma_D} \propto 1/r^2$,
while for $r \stackrel{<}{\sim} r_D$, $F_{\gamma_D}$ 
falls much more modestly. 
\footnote{
This behavior assumes that the halo is (at least approximately) optically thin. 
That is, the optical depth, $\tau$, along the dark photon's path between
its production point and absorption point satisfies $\tau \stackrel{<}{\sim} 1$.
In general $\tau$ depends on the photon's frequency (energy) so that
in practice the halo might only be optically
thin for a range of dark photon frequencies.  Actually this detail may not be so important as
the energy transported to the halo is dominated by
dark photons with optical
depths $\tau \stackrel{<}{\sim} 1$. 
This conclusion is also supported by numerical work \cite{foot2,foot3} which includes the effects of
finite optical depth and allows for a  range of possible shapes for the dark photon frequency spectrum
generated by supernovae.} 
In fact, simple calculations \cite{foot4} 
suggest that $F_{\gamma_D} \propto log(r)$ for $r \stackrel{<}{\sim} r_D$. 
These features 
motivate a quasi-isothermal dark matter distribution, with dark matter 
core radius, $r_0$, scaling with $r_D$, $r_0 \sim r_D$.

The dark matter density profile, Eq.(\ref{2z}), scales with the dark photon energy flux. This also suggests 
a correlation with the dark photon source, the type II supernovae and by extension, the local star formation
rate.
This logic suggests that the dark matter properties, in particular the density profile, should be closely
linked with the baryonic properties. 
The $r_0 \sim r_D$ scaling discussed above is one example of this.
Another example is
a correlation between the supernova rate and asymptotic value of the rotation velocity,
which has been shown \cite{foot4} to be equivalent to the Tully-Fisher relation \cite{tf}. Eq.(\ref{2z}) also
holds locally:
Regions of a galaxy with low gas density can have reduced supernova 
rates leading to a locally depressed dark photon energy flux, $F_{\gamma_D}$. Halo dynamics should also
adjust the cooling rate in a corresponding fashion, thereby necessitating a lower dark matter density
in that region.
Thus galaxies with wiggles and other features in their baryonic gas distribution can have 
corresponding features in the dark matter distribution, and hence leave an imprint on the rotation curve.
The purpose of this paper is to explore this possibility. To do so, we utilize a very simple
formula for the dark matter density distribution, suggested by the balancing of the dark
plasma cooling and heating processes, described above. We then study the implied rotation
curves and examine specific examples. 

The outline of this paper, then, is as follows. In section 2 we give some of the particle 
physics details of a candidate dissipative dark matter model.
In section 3, we develop the necessary tools to describe the dark matter density
in terms of baryonic properties in this dissipative dark matter approach.
In section 4 we examine some general features of the predicted dark matter density 
and rotation curves for generic disk galaxies. In section 5 we apply this formalism 
to some specific examples and 
in section 6 we conclude.

\section{Dissipative dark matter models}

\vskip 0.4cm

The idea that dark matter originates from a hidden (or dark) sector 
weakly coupled to the standard model
has become quite widely discussed in recent years, e.g.
\cite{hid0,footoldz,hid1,hidb,hid2,hid3,hidz,hid4,hid5} 
(see also \cite{footreview} for a more detailed bibliography).
Such models allow for a very rich dark matter phenomenology, especially if the 
hidden sector contains unbroken gauge interactions.
If the hidden sector features stable matter particles charged under an unbroken $U(1)'$ gauge symmetry,
then dark matter properties can become extremely interesting. In particular dissipative self interactions
causing dark matter to cool via the emission of massless dark photons become possible.
Although the dark sector can be quite strongly self interacting, dark sector interactions with electrons, 
protons, and other ordinary sector
particles, are relatively weak. In the model considered here, only gravity
and a very small photon - dark photon kinetic mixing interaction (to be
defined in a moment) are presumed. 

A very generic dissipative dark matter model was considered in \cite{foot4} 
and shown to provide a consistent explanation for the CMB anisotropy spectrum and large scale structure. 
In addition it was argued that the same model could also potentially explain small scale structure:
the cored dark matter distribution in galaxies, Tully-Fisher and other scaling relations etc.
We refer the reader to that reference for these details, here we just review a few essential  
elements of that model to illustrate the basic ideas.

Consider then a 
hidden sector featuring two massive stable particles, $F_1$ and $F_2$. 
These dark matter particles can be either
bosonic or fermionic; in the following we will assume they are spin $1/2$ fermions for definiteness.
These particles are presumed to be charged under an unbroken $U(1)'$ gauge symmetry, that is they
interact with a massless $U(1)'$ gauge boson -  the dark photon.
The particle interactions can then be described by the Lagrangian:
\begin{eqnarray}
{\cal L} = {\cal L}_{SM} + {\cal L}_{dark} + {\cal L}_{int}
\label{l1}
\end{eqnarray}
where ${\cal L}_{SM}$ is the standard model Lagrangian, and ${\cal L}_{dark}$ describes
the interactions of the dark sector:
\begin{eqnarray}
{\cal L}_{dark} = 
 -\frac{1}{4}F ^{'\mu \nu} F_{\mu \nu}^{'} + 
\overline{F}_1(iD_{\mu}\gamma^{\mu} - m _{F_1})F _1 +
\overline{F}_2(iD_{\mu}\gamma^{\mu} - m _{F_2})F _2 
\label{dark}
\ .
\end{eqnarray}
Here, $F_{\mu \nu}^{'} = \partial_{\mu} A_{\nu}^{'} - \partial_{\nu} A_{\mu}^{'}$ 
is the field-strength tensor 
associated with the $U(1)^{'}$ gauge interaction, $A _{\mu}^{'}$ 
being the relevant gauge field. 
Also, the covariant derivative is $D_{\mu} \equiv \partial_{\mu} + ig^{'}Q^{'} A _{\mu}^{'}$, 
where $g^{'}$ is the coupling constant relevant to this gauge interaction.
The dark sector may have much more complexity, additional gauge interactions, additional particles etc,
but only the low energy effective theory will be important 
for the astrophysical implications discussed in this paper. This means that the QED-like Lagrangian, Eq.(\ref{dark}),
can actually represent a wide class of hidden sector theories.

In such a theory it is natural to expect dark matter in the Universe to be the result of a particle - antiparticle asymmetry,
as the symmetric component would be 
expected to efficiently annihilate in the early Universe (at least for much of the parameter space of interest).
That is, dark matter in the Universe is composed of $F_1$ and $F_2$ particles (with negligible proportion
of antiparticles).
Dark matter asymmetry and local
neutrality of the Universe then imply:
\begin{eqnarray}
n _{F_1}Q _{F_1}^{'} + n _{F_2}Q _{F_2}^{'} = 0 
\end{eqnarray}
where $n _{F_1}$ and $n _{F_2}$ are the number densities of $F _1$ and
$F _2$ respectively. Evidently the $U(1)'$ charges of $F_1$ and $F_2$, $Q_{F_1}^{'}$ and $Q_{F_2}^{'}$,
must be opposite in sign, but not necessarily equal in
magnitude. This is, of course, quite analogous to the
situation with ordinary matter ($F_1 \sim$ electron, $F_2 \sim$ proton).

The ${\cal L}_{int}$ part in Eq.(\ref{l1}) has yet to be defined.
Constraints from gauge invariance and renormalizability strongly constrain possible interactions connecting
ordinary and hidden sector particles. In this particular model, these constraints limit ${\cal L}_{int}$
to just the kinetic mixing interaction \cite{foothe}, which leads to photon - dark photon kinetic mixing:
\begin{eqnarray}
{\cal L}_{int} = - \frac{\epsilon}{2}F^{\mu \nu} F_{\mu \nu}^{'} \ .
\end{eqnarray}
This kinetic mixing interaction effectively provides the dark fermions, $F_1$ and $F_2$,
with a tiny ordinary electric charge proportional to the dimensionless parameter $\epsilon$ \cite{holdom}.

A tiny kinetic mixing interaction allows ordinary type II supernovae to 
become a huge energy source for the dark sector.
Recall that in
standard theory nearly all of a supernova's core collapse energy, typically around $few \times 10^{53}$ ergs,
is converted into neutrinos in the hot core on a time scale of about 10 seconds.
The dozen or so neutrino events detected following SN 1987A supports this
basic picture \cite{kam,kam2}. However there are still substantial uncertainties which means that 
the energy carried away by exotic particles can be up to
around $50\%$ of the total core collapse energy \cite{raf}.
It turns out that for
kinetic mixing strength $\epsilon \sim 10^{-9}$
around half of a supernova's core collapse energy ($\sim 10^{53}$ ergs)
can be converted into  light hidden sector particles, say $F_1$, if their mass is less than the 
supernova core temperature, $\sim$ 30 MeV \cite{raf}. 
\footnote{Natural units with $\hbar = c = k _B = 1$ are assumed.}
[The $F_2$ dark matter particle can be much heavier, with GeV-TeV scale preferred \cite{foot4}.]
The relevant particle processes include plasmon decay  where the effectively massive photon propagating
in the plasma decays into $\bar F_1 F_1$ pairs.
Further processes, such as $\bar F_1 F_1 \to \gamma_D \gamma_D$, $F_1 F_1 \to F_1 F_1 \gamma_D$ etc,
convert this energy into dark photons in the region around the supernova, which ultimately
escape and provide the heat source for the dark halo.
This huge heat source, averaging to around $\sim 10^{44}$ ergs per second for the Milky Way (for $\epsilon \sim 10^{-9}$), 
can potentially 
compensate for the energy lost due to dissipative interactions.
In fact, this imbues the dark halo with very rich and nontrival dynamics. 
The halo can contract or expand, and it is anticipated that the halo evolves until a steady-state configuration
is reached where the energy going into the halo sourced from supernovae,
as described above,
approximately balances the
energy dissipated.
This is the key assumption going into the dissipative dark matter scenario advocated here.

The above particle physics model provides a simple candidate for dissipative dark matter.
Importantly, the analysis of the present paper is, in many respects, model independent - applicable it 
is hoped to many possible dissipative dark matter models. Among these
is the  mirror dark matter model,
where the hidden sector is an exact duplicate of the standard model. That allows an exact, unbroken
$Z_2$ discrete symmetry to exist which swaps each ordinary matter with a mass degenerate `mirror' partner \cite{flv}.
The matter particles in this case are the mirror electron of mass $0.511$ MeV and mirror nuclei, $H', He', O', Fe',...$
and the role of the dark photon is played by the mirror photon \cite{footreview,footoldz}.
Dissipative dark matter
models with light rather than massless dark photons, or even with light scalar particle(s),
might also be possible.

\vskip 0.2cm

\section{Formalism}

\vskip 0.5cm

The physical picture for disk galaxies 
is that they are embedded 
in a galactic halo that takes the form of a plasma 
\footnote{There may also be a subcomponent of dark stars. However, the total mass (and distribution)
of that subcomponent is
uncertain and assumed negligible for the purposes of the present analysis.}
composed of dark matter particles \cite{footvolkas}.
\footnote{Elliptical and dwarf spheroidal galaxies form quite a different class,
as these galaxy types are largely devoid 
of baryonic gas and show little current star formation activity. The curious reader is referred to \cite{foot4}
for a discussion on how these galaxies might fit into this dissipative dark matter framework.}
This theme has been developed over the last 
few years in the context of mirror dark matter \cite{foot1,foot2,foot3,footsil} and the simple
dissipative model of the previous section  \cite{foot4}. Here we build on this recent work, with the 
aim of describing such dissipative dark matter in a fairly model independent manner.

In the context of the specific hidden sector model described in the previous section,
the dark matter plasma dissipates energy via various processes which lead to dark photon emission, such
as dark bremsstrahlung.
The halo is also subject to heating, sourced by kinetic mixing induced processes in
the core of type II supernovae in the disk.
These processes 
($\gamma \to \bar F_1 F_1$, $\bar e e \to \bar F_1 F_1$, $\bar e e \to \bar F_1 F_1 \gamma_D$ 
etc) 
convert up to around half of a supernova's core collapse energy 
into light hidden sector particles.
Ultimately this energy is transformed into dark photons in the region around the supernova. 
These dark photons then transport this energy to the halo, where it can be absorbed via dark photoionization.

Using spherical coordinates and
setting the presumed thin disk at $\theta = \pi/2$, the energy flux of dark photons at a point $P = (r,\theta,\phi)$ 
within an optically thin halo is given by:
\begin{eqnarray}
F_{\gamma_D} (r,\theta,\phi) =  
\int
d\widetilde{\phi} \int d\widetilde{r} \ \widetilde{r} 
\ \frac{\Sigma
({\widetilde{r},\widetilde{\phi}})}{4\pi[r^2 + {\widetilde{r}}^2 - 2r\widetilde{r}  \sin\theta \cos
(\widetilde{\phi}-\phi)]}
\ .
\label{1}
\end{eqnarray}
Here, $\Sigma (\widetilde{r},\widetilde{\phi})$
is the luminosity of dark photons per unit area produced on the disk, assumed to originate from ordinary type II supernovae.
Since supernovae are discrete events, this quantity represents an average over a reasonable time period (say last 10 million
years).

The flux of dark photons will heat the dark halo. The halo also cools via dark bremsstrahlung
and other processes. The key assumption is that the halo dynamically adjusts (contracts or expands)
so that the heating and cooling rates balance at any particular point, $P$.
Given that the heating rate is proportional to 
the product of the dark matter density and dark photon energy flux and the cooling rate is proportional to 
this density squared it follows that $\rho \propto F_{\gamma_D}$ [Eq.(\ref{1z}) and Eq.(\ref{2z})].
That is, the dark matter mass density has the form:
\begin{eqnarray}
\rho(r,\theta,\phi) =  
\lambda \int
d\widetilde{\phi} \int d\widetilde{r} \ \widetilde{r} 
\ \frac{\Sigma_{SN}
({\widetilde{r},\widetilde{\phi}})} {4\pi[r^2 + {\widetilde{r}}^2 - 2r\widetilde{r}  \sin\theta \cos
(\widetilde{\phi}-\phi)]}
\label{3z}
\end{eqnarray}
where $\lambda$ is a constant which depends on the cross section, supernovae dark photon energy spectrum etc.
In the integrand above, $\Sigma_{SN}$ is the type II supernova rate in the disk 
(the proportionality constant in replacing the
dark photon luminosity in the integrand of Eq.(\ref{1}) with $\Sigma_{SN}$ is also absorbed into $\lambda$).
In principle $\lambda$ also depends on the halo temperature at the point $P$. However, 
we will assume, for simplicity, an isothermal
halo so that the constant $\lambda$ is independent of halo location, i.e. $r, \theta, \phi$.
[Of course, $\lambda$ may well vary somewhat between different galaxies.]

The above equation gives a very simple description of the dark matter  
distribution in galaxies. 
Furthermore, it is intimately connected with the distribution of ordinary baryons through
the type II supernova rate per unit area of the disk, $\Sigma_{SN} (\widetilde{r},\widetilde{\phi})$.
The previous studies \cite{foot2,foot3,foot4} assumed that $\Sigma_{SN} (\widetilde{r},\widetilde{\phi})$ could be
approximated by the stellar distribution, taken to be a Freeman disk \cite{freeman}. This could only be a fairly crude
approximation as the Freeman disk is a kind of average over the past history of the galaxy; the quantity we 
require is the supernovae rate at the present epoch. Since supernovae are the final evolutionary stages of large
stars which have a short lifespan (typically less than 10 million years) the rate $\Sigma_{SN}(\widetilde{r},\widetilde{\phi})$
should be correlated with the current star formation rate in the galaxy under consideration.

Under the simplifying assumption of an azimuthally symmetric disk, the dark matter density, $\rho$ [Eq.(\ref{3z})], is also
azimuthally symmetric.
The contribution to the gravitational 
acceleration at a point in the plane of the disk due to the dark matter distribution, 
Eq.(\ref{3z}),  therefore depends only on the distance $r$ and is given by
\begin{eqnarray}
g(r) = G_N \int d\widetilde{\phi} \int d\cos\widetilde{\theta} 
\int d\widetilde{r} \ \widetilde{r}^2 \ {\rho (\widetilde{r},\widetilde{\theta}) \cos\omega \over 
r^2 + {\widetilde{r}}^2 - 2r \widetilde{r}  \sin\widetilde{\theta} \cos
\widetilde{\phi}}
\ .
\label{4z}
\end{eqnarray}
Here $G_N$ is Newton's constant and
\begin{eqnarray}
\cos\omega = {r - \widetilde{r}\sin\widetilde{\theta}\cos\widetilde{\phi}
\over \sqrt{r^2 + {\widetilde{r}}^2 - 2r \widetilde{r}  \sin\widetilde{\theta} \cos
\widetilde{\phi}}}
\ .
\end{eqnarray}
In this azimuthally symmetric case, the motion of the baryons in the disk can be assumed to be 
circular
and the dark matter halo contribution to the rotational velocity is then given by $v_{halo}^2/r = g(r)$.

If the halo does indeed evolve to a steady-state configuration where heating and cooling
rates locally balance then the distribution of the dark matter out to
some distance, $r_{halo}$, should be fixed by Eq.(\ref{3z}).
This means that for $0 < r \stackrel{<}{\sim} r_{halo}$ a kind of equilibrium configuration ensues
so that the dark matter distribution
is approximately independent of the initial conditions and early history of the galaxy. 
\footnote{
Regarding the early history of the halo evolution,
we expect that at very early times ($t \stackrel{<}{\sim} Gyr$)
the dark matter that seeded the galactic halo
would have collapsed, possibly into a disk.
Shortly thereafter, 
following the collapse of ordinary matter and consequent star formation activity, the dark matter
was reheated. 
This reheating expands the dark matter
gas component
into the plasma halo and eventually the system should relax into a steady-state configuration
where the heating and cooling rates locally balance. This is the presumed state of disk galaxies at the current
epoch. See also \cite{footreview} for some more detailed discussion.}
In particular, the distribution is also independent of the total dark matter mass of the halo, $m_{dark}$.
However,
the size of $r_{halo}$ is expected to depend on this quantity: 
\begin{eqnarray}
m_{dark} = \int \rho dV \sim 4\pi \rho_0 r_0^2 r_{halo}
\end{eqnarray}
where we have replaced $\rho$ with the quasi-isothermal distribution, an approximation roughly valid, 
especially at large distances, $r \gg r_D$ (as will be seen in section 4).
Observations reveal that rotation curves are generally flat at the largest observed distances in a given
galaxy, which indicates that $r_{halo}$ is typically greater than this observed edge of the galaxy.

\section{Generic Galaxies}

We now examine the case of a `generic' disk galaxy. Specifically we consider
a thin disk galaxy (with no bulge) with stellar mass distribution described by a Freeman
disk \cite{freeman}:
\begin{eqnarray}
\Sigma^* (r) = \frac{m _D}{2\pi r_D ^2}e ^{-\frac{r}{r_D}} 
\label{z1}
\end{eqnarray}
where $m_D$ is the stellar mass of the disk and $r_D$ is the disk scale length.
To solve Eq.(\ref{3z}) we require the 
type II supernova rate in the disk, $\Sigma_{SN} (r,\phi)$.
Here we shall assume
azimuthal symmetry and take $\Sigma_{SN} (r) \propto \Sigma^*(r)$. 
Of course this can only be a very crude approximation, but can nevertheless provide
useful  insight for generic disk galaxies.
With this assumption, the type II supernova rate per unit area is then:
\begin{eqnarray}
\Sigma_{SN} (r) = {\Sigma^* (r) \over m_D} R_{SN}
\label{z2}
\end{eqnarray}
where $R_{SN}$ is the total supernova rate of the generic galaxy under consideration.
We further assume that this total supernova rate scales with stellar mass via: 
$R_{SN} \propto \sqrt{m_D}$, which is consistent with the supernova study \cite{sn}.

With these baryonic relations,
it is straightforward to numerically
solve Eqs.(\ref{3z},\ref{4z}) to obtain the dark halo density distribution and rotation curve.
In doing so, we assume that the coefficient $\lambda$ is a constant, which we set by fixing 
the maximum rotational velocity of a $m_D = 10^{11} m_\odot$ generic galaxy to be 250 km/s (a typical value for such a galaxy).
\footnote{In realistic models we expect some variation of $\lambda$ between galaxies. In fact, $\lambda \propto 1/\sqrt{T_{halo}}$
follows in models where dark bremsstrahlung is the dominant cooling process.
Simple analytic calculations suggest that the halo temperature is related to the asymptotic rotational velocity via
$T_{halo} \sim \bar m v_{rot}^2/2$ where $\bar m$ is the mean mass of the dark matter particles forming the plasma \cite{foot1}. 
Utilizing the Tully-Fisher relation we then have $\lambda \propto (L_B)^{-1/4}$, where $L_B$ is the B-band luminosity
of the galaxy in question.}

Consider first the resulting dark matter distribution, Eq.(\ref{3z}).
This distribution is 
not spherically symmetric, due to the spherically asymmetric heating
from the disk.  In figure 1 we plot $\rho (r,\theta)$
for $\theta = 0$ (direction normal to the disk), 
$\theta = \pi/6$, $\theta = \pi/4$, $\theta = \pi/3$ and $\theta = 2\pi/5$.
The density has a log divergence as $\theta \to \pi/2$ (plane of the disk).
As the figure shows, the dark matter halo approaches spherical symmetry for $r \gg r_D$,
with significant departures  from spherical symmetry in the inner region ($r \stackrel{<}{\sim} r_D$).

\vskip 0.5cm
\centerline{\epsfig{file=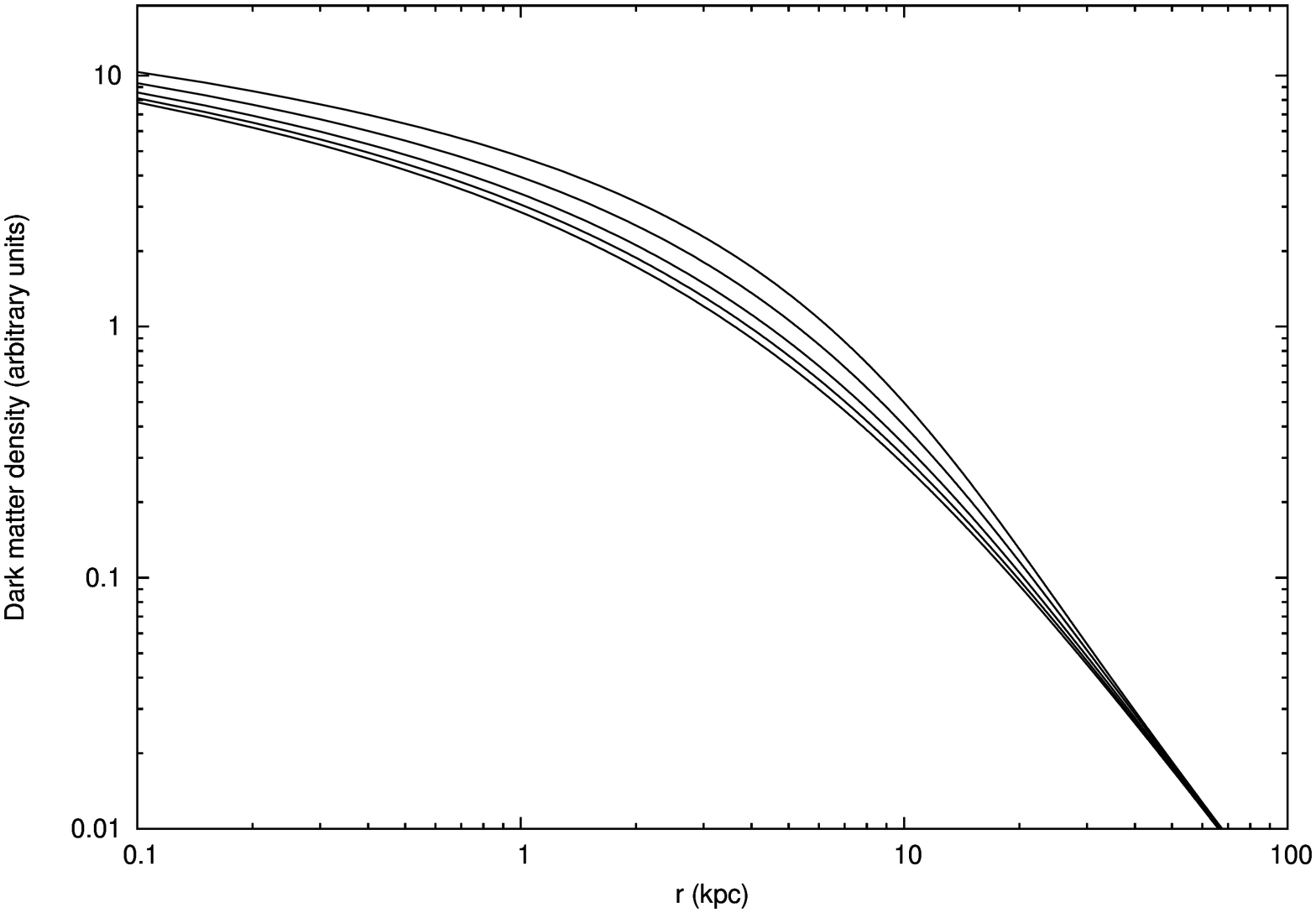,angle=270,width=11.7cm}}
\vskip 0.3cm
\noindent
{\small Fig. 1: Dark matter density, $\rho(r,\theta)$ derived from dissipative dynamics  
[Eqs.(\ref{3z},\ref{z2})] for various $\theta$ values. 
Curves from bottom to top are for:
$\theta = 0$ (direction normal to the disk), $\theta = \pi/6$, $\theta = \pi/4$, $\theta = \pi/3$ and $\theta = 2\pi/5$.
The reference values $m_D = 10^{11} m_\odot, \ r_D = 4.3$ kpc are assumed.}

\vskip 0.6cm
\centerline{\epsfig{file=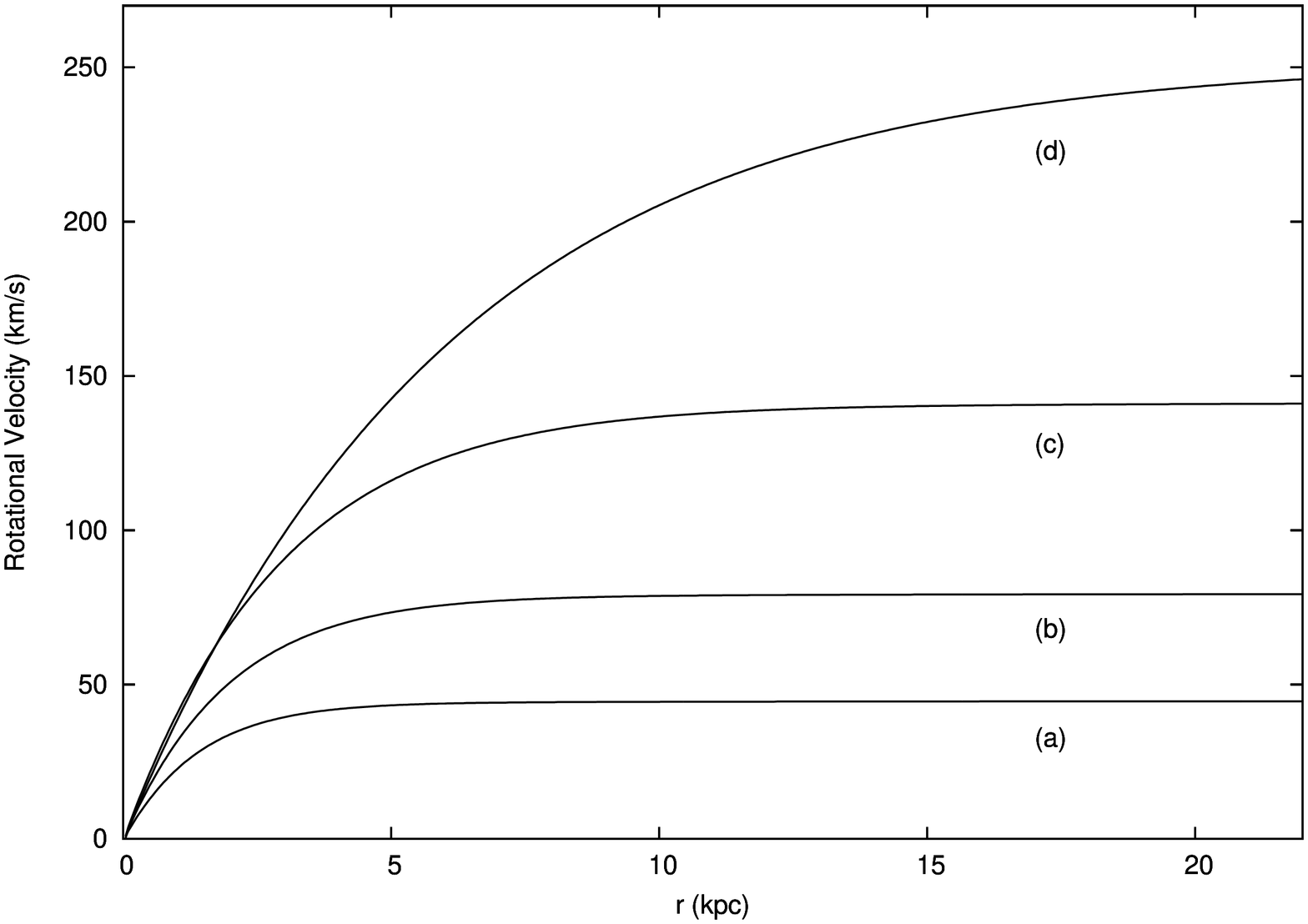,angle=270,width=11.7cm}}
\vskip 0.3cm
\noindent
{\small Fig. 2: Generic rotation curves predicted from this dynamics, Eqs.(\ref{3z},\ref{4z},\ref{z2}). The four curves are for
parameters: (a) $m_D = 10^{8} m_\odot, \ r_D = 1.0$ kpc, (b) $m_D = 10^{9} m_\odot, \ r_D = 1.4$ kpc, 
(c) $m_D = 10^{10} m_\odot, r_D = 2.1$ kpc, 
and (d) $m_D = 10^{11} m_\odot, \ r_D = 4.3$ kpc.
The curves represent only the dark matter contribution to the rotational velocity.}

\vskip 1.2cm
We now turn to the rotation curves, Eq.(\ref{4z}).
Figure 2 shows the result of
numerically solving this equation for
four illustrative examples with stellar masses of $m_D = 10^{8} m_\odot$, $m_D = 10^{9} m_\odot$, $m_D = 10^{10} m_\odot$ and 
$m_D = 10^{11} m_\odot$. The corresponding values for the disk scale length (given in the figure caption) are  
typical of high surface brightness disk galaxies.

In figure 3 we compare the predicted rotation curves (dark matter contribution only) 
for two galaxies with the same baryonic mass $m_D = 10^{10} m_\odot$ but with
two different disk scale lengths: (a) $r_D = 2$ kpc and (b) $r_D = 7$ kpc. 
These parameters correspond roughly to the
high surface brightness galaxy NGC 2403 (a) and low surface brightness galaxy UGC 128 (b).
[To do an actual fit to the rotation curve data for these two galaxies, a more sophisticated 
treatment of $\Sigma_{SN}$ is recommended, such as 
the approach in section 5 which makes use of a
Kennicutt-Schmidt-type relation.]
Although both galaxies have the same asymptotic value for the rotational velocity,
the low surface brightness galaxy has a much shallower rotation curve.
This behavior has been known for many years, see in particular the discussion in \cite{ugc,stacy}, 
and is very simply explained
with dissipative dark matter dynamics.

\vskip 0.8cm
\centerline{\epsfig{file=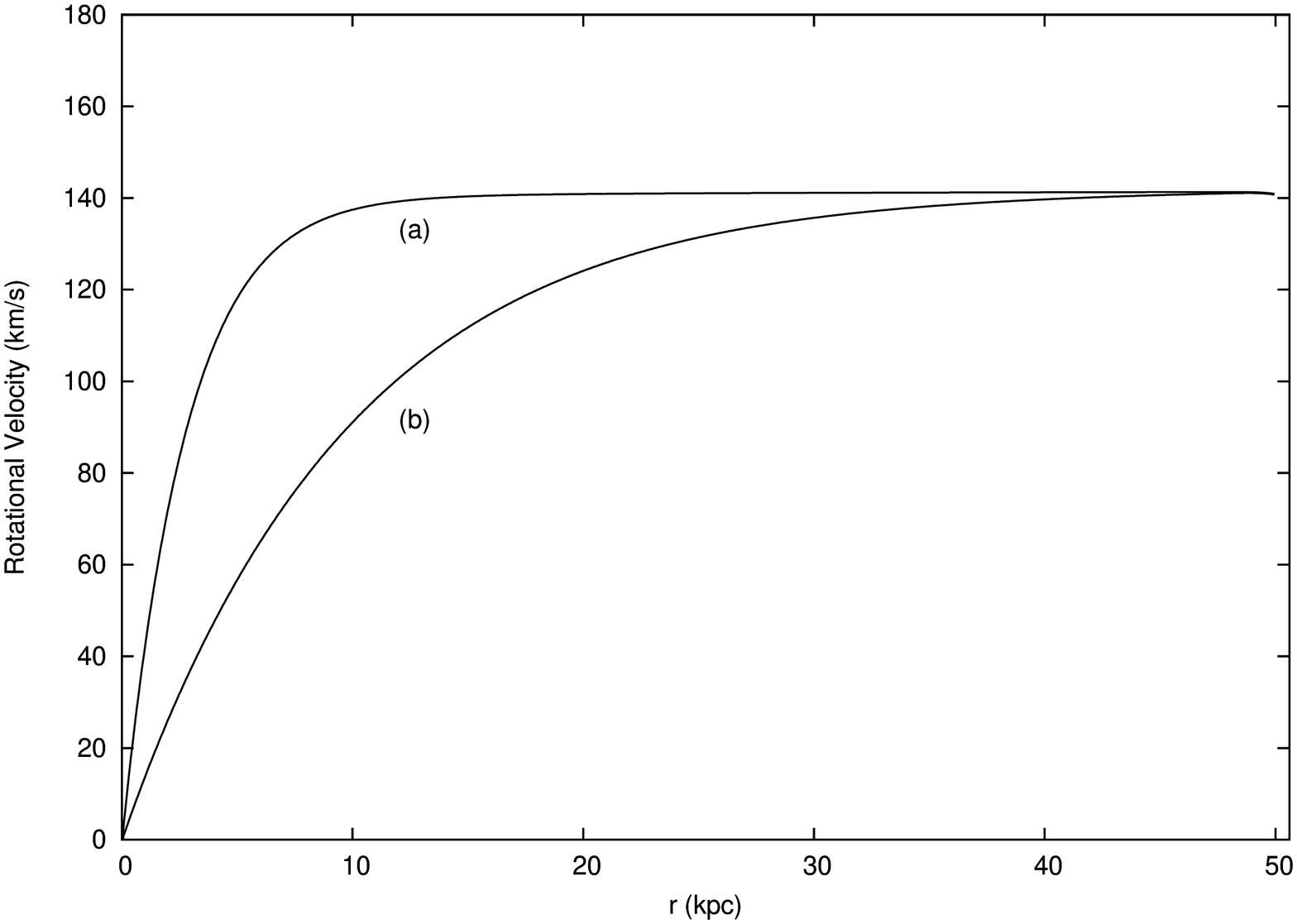,angle=270,width=11.7cm}}
\vskip 0.3cm
\noindent
{\small Fig. 3a: Predicted rotation curve (dark matter only contribution) for
two galaxies with the same baryonic mass $m_D = 10^{10} m_\odot$ but with
two different disk scale lengths: (a) $r_D = 2$ kpc and (b) $r_D = 7$ kpc.} 

\vskip 1.0cm
\centerline{\epsfig{file=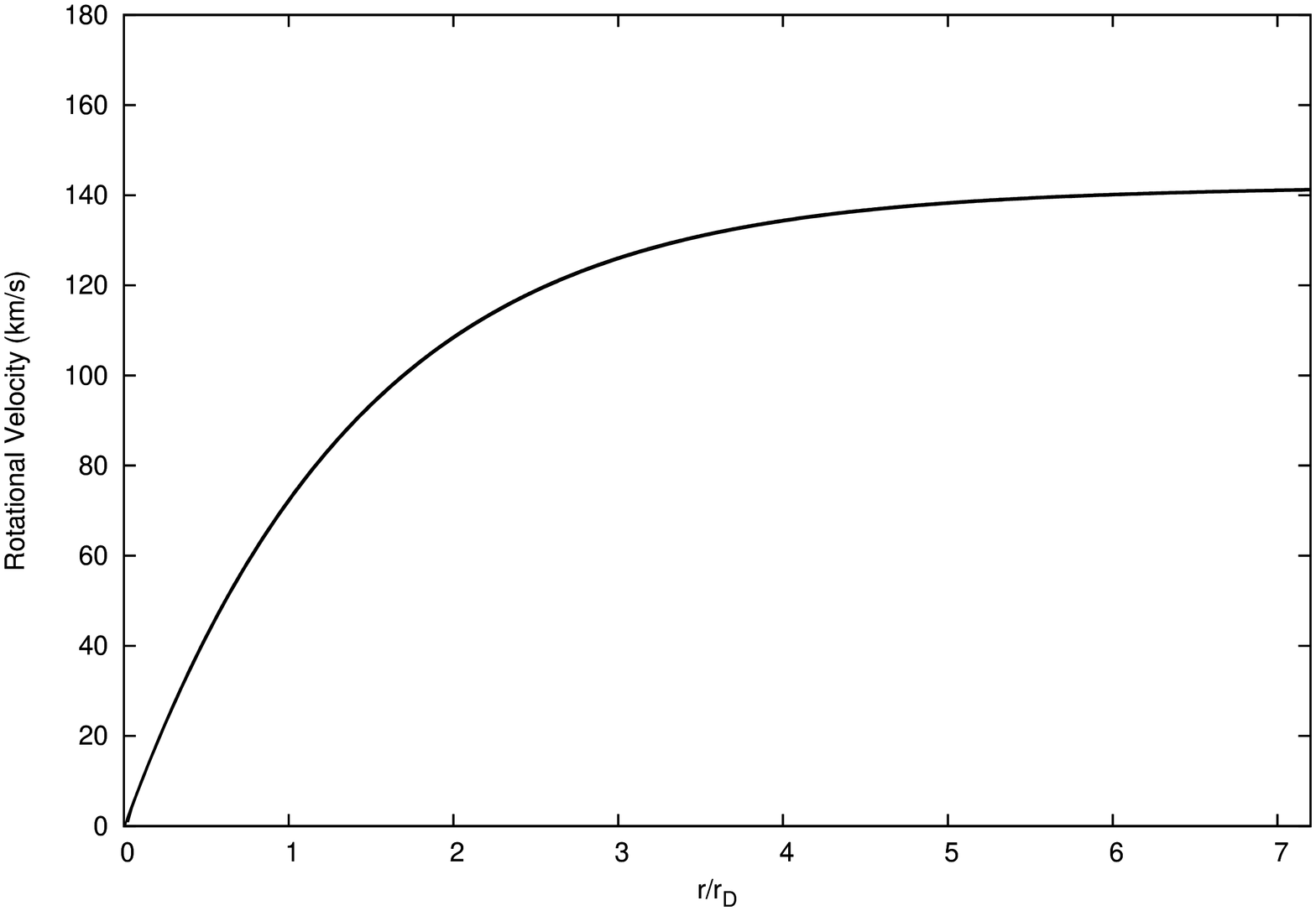,angle=270,width=11.7cm}}
\vskip 0.3cm
\noindent
{\small Fig. 3b: The same two curves of figure 3a plotted in terms of the dimensionless
variable $r/r_D$.
The curves coincide as $r_D$ is the only length scale in the problem. 
}

\vskip 1.1cm
\newpage

Phenomenologically, it has long been observed that spherically symmetric cored dark matter distributions 
provide a reasonable fit to the rotation curves of typical disk galaxies, e.g. \cite{blok,rubin2,blok2,salucci}.
In this context, the spherically symmetric quasi-isothermal distribution:
\begin{eqnarray}
\rho_{ISO} (r) = {\rho_0 r_0^2 \over r^2 + r_0^2}
\ 
\label{iso}
\end{eqnarray}
with rotation velocity
\begin{eqnarray}
{v_{rot}^2 \over r} = {G_N \over r^2} \int_0^r \ \rho_{ISO} (\widetilde{r}) \ 4\pi
\widetilde{r}^2 d\widetilde{r}
\label{iso2}
\end{eqnarray}
is often considered. Other closely related cored profiles have also been discussed
in the literature, such as the Burkert profile \cite{burk,burk2}.

Although the dark matter density profile resulting from the assumed dissipative dynamics is not spherically symmetric,
it turns out that the dark halo contribution to the rotational velocity can be approximated by
the above spherically symmetric quasi-isothermal distribution.
We illustrate this in
figure 4, which compares the generic rotation curves of figure 2 [derived from Eq.(\ref{3z}) and Eq.(\ref{z2})] 
with the rotation curves resulting from the spherically symmetric quasi-isothermal distribution, Eq.(\ref{iso2}).
Furthermore, the approximate agreement of the two curves requires
$r_0 \approx r_D$ and $\rho_0 \approx 7\times 10^{-2}\ m_\odot/pc^3$, independently of $m_D$ over the 
considered range: $10^8 < m_D/m_\odot < 10^{11}$.
This value for $r_0$ is somewhat below best fit values, which prefer $r_0 \approx 2r_D$ \cite{donato1}.
We suspect that this is due, at least in part, to our use of $\Sigma^*$ to model $\Sigma_{SN}$.
In the next section we will use the Kennicutt-Schmidt-type relation which we find
does a reasonable job in explaining rotation curves.

Also, the result that $\rho_0$ was found to be roughly constant  
relied on $\lambda$ being a galaxy independent quantity and also the  
assumed $R_{SN} \propto \sqrt{m_D}$ scaling.
In models where dark bremsstrahlung dominates the halo cooling,
$\lambda \propto 1/\sqrt{T_{halo}}$, where $T_{halo}$ is the temperature of the dark matter
halo. Simple analytic calculations as in \cite{foot4}
suggest $T_{halo} \propto \rho_0 r_0^2$. If this is taken into account, then we would have $\rho_0 \propto 1/r_0^{2/3}$,
a galactic scaling relation that is roughly consistent with observations \cite{korm,donato} (the Tully-Fisher
relation \cite{tf} also follows from similar arguments \cite{foot4}).
These scaling relations, though,  are limited by uncertainties such as that of the 
$R_{SN} \propto \sqrt{m_D}$ scaling.

\vskip 0.1cm
\centerline{\epsfig{file=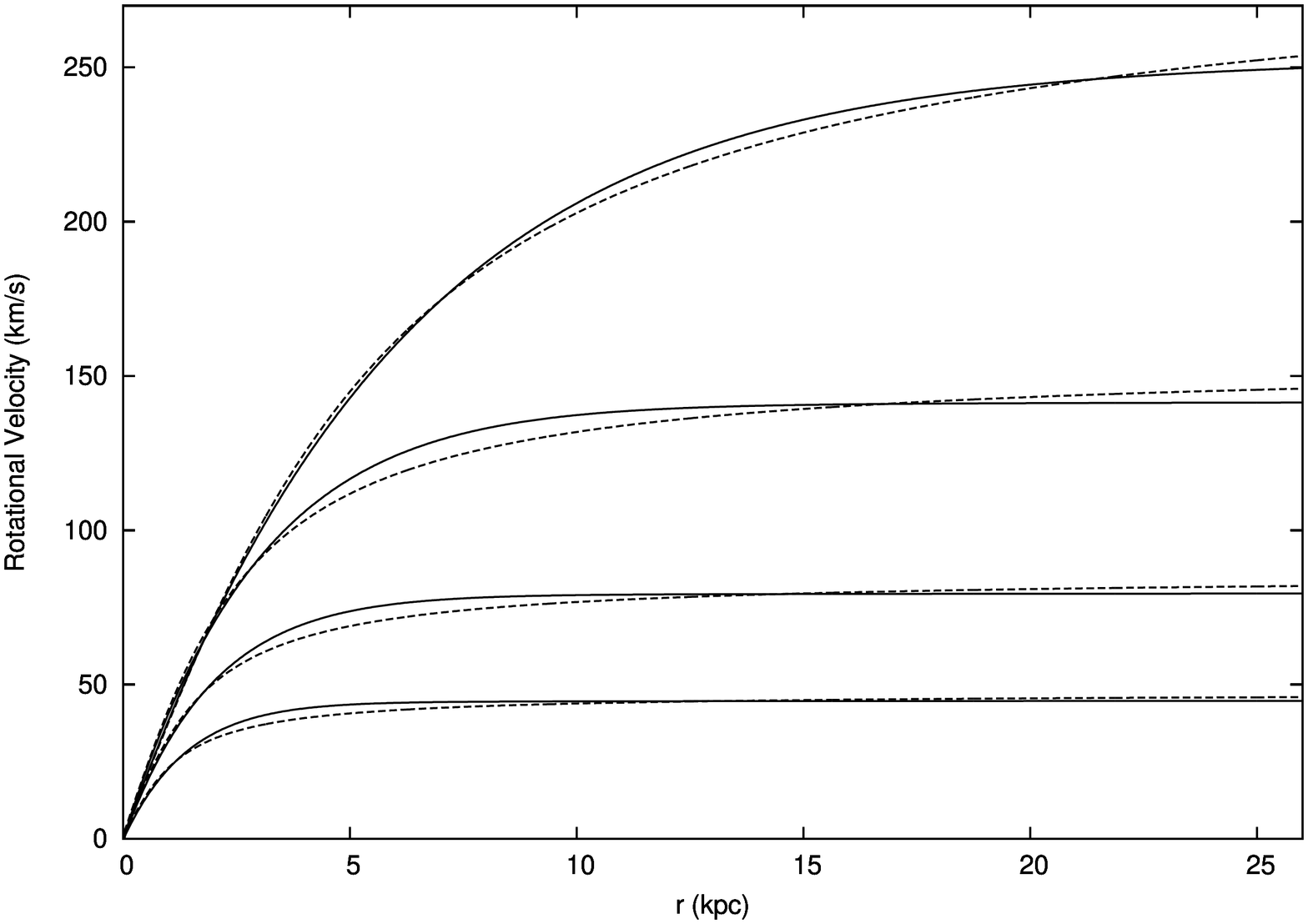,angle=270,width=11.7cm}}
\vskip 0.5cm
\noindent
{\small Fig. 4: Comparison between the rotation curves from 
dissipative dynamics for the examples in figure 2 (sold lines) with that using the quasi-isothermal profile [Eq.(\ref{iso})]
(dashed lines).  
}


\vskip 0.8cm

\section{Specific examples}
\vskip 0.3cm

Over the last few years, high quality rotation curve data has become available from surveys of nearby
galaxies, e.g. \cite{things,oh}. 
Here we shall examine examples contained in the LITTLE THINGS survey of 26 nearby dwarf galaxies \cite{oh}. Additionally,
we consider also 
the small spiral galaxy, NGC 1560 \cite{bor,gen}.  The selected sample mostly feature flat or rising rotation 
curves out to the largest measured distance. 

The study \cite{oh} compared the rotation curves of the LITTLE THINGS sample with the quasi-isothermal 
profile, Eq.(\ref{iso}) and also the NFW profile \cite{nfw}
\begin{eqnarray}
\rho_{NFW}(r) = {\rho_i \over (r/r_s)(1 + r/r_s)^2}
\ . 
\label{nfw99}
\end{eqnarray}
It was found that the NFW profile generally gave very poor fits to the observed rotation curves,
while the quasi-isothermal profile generally fared much better.
As discussed in the previous section, the quasi-isothermal profile was historically introduced purely phenomenologically,
without reference to any dark matter theory, 
but nevertheless can be viewed as a zeroth order approximation to dissipative dark matter (figure 4). 
Note though, that
dissipative dark matter provides an even more constrained description since the core scale length ($r_0$)
correlates with the disk scale length $r_D$, via the association of the heat sources with core collapse supernovae
in the disk. The purpose of the current paper is to illustrate these ideas with specific examples, and 
postpone a more detailed study to future work \cite{prep}.

As in section 4 we shall derive the rotation curves using Eqs.(\ref{3z},\ref{4z}) for the dark matter contribution. 
However, instead of using $\Sigma^*$ for the supernova rate in the disk,
we shall use a Kennicutt-Schmidt-type relation: $\Sigma_{SN}(\widetilde{r})   \propto [\Sigma_{gas}(\widetilde{r})]^N$.
Such a relation allows $\Sigma_{SN}$ to be related to local properties of each galaxy.
Explicitly, we use
\begin{eqnarray}
\rho(r,\theta) =  
\lambda' \int
d\widetilde{\phi} \int d\widetilde{r} \ \widetilde{r} 
\ \frac{[\Sigma_{gas}
({\widetilde{r}})]^N} {4\pi[r^2 + {\widetilde{r}}^2 - 2r\widetilde{r}  \sin\theta \cos
\widetilde{\phi}]}
\label{9z}
\end{eqnarray}
where the coefficient is now $\lambda'$, having  absorbed the proportionality constant
of the Kennicutt-Schmidt relation into its definition.
For $N=2$, $\lambda'$ has units $pc/m_\odot$.

For the exponent, $N$, in the Kennicutt-Schmidt type relation, 
we initially considered the range: $1.0 < N < 2.0$, but for the sample of galaxies considered 
here it was found that the rotation curve data
could be reasonably well explained if $N = 2.0$ (although lower values, e.g. $N = 1.5$ also
gave reasonable results).
The value of $N = 2.0$ is consistent with the star formation rate found by Schmidt \cite{schmidt},
who considered star formation rates in the Milky Way,
but somewhat higher than the value obtained by the more recent study of Kennicutt, $N \approx 1.4$ \cite{ken}
(see also the recent review \cite{ken2}).
However the Kennicutt study considered the total star formation rate in units $m_{\odot} \ yr^{-1}$ averaged
over the disk for a given galaxy,
while the quantity of interest in the present context is the {\it local} formation rate of large stars, 
$m_{star} \stackrel{>}{\sim} 8 m_\odot$,
the progenitors of type II supernovae.
Anyway, with the exponent fixed at $N=2.0$,
we could attempt to fit each galaxy with just one parameter, $\lambda'$.
[Although we allow $\lambda'$ to vary between different galaxies, it is found that there is only a modest
variation, within
a factor of two or so between the 10 galaxies considered, with central value around $\lambda' \approx 0.001 \
pc/m_\odot$.]

We first consider the small spiral galaxy, NGC 1560 \cite{bor,gen}.
This example is particularly interesting because it shows a small `wiggle' in the rotation curve, at
around $r \approx 5$ kpc. 
The gas density for this galaxy is given in \cite{gen}, reproduced in figure 5a,  
also has this feature at $r \approx 5$ kpc.
Although the wiggle is quite small in the azimuthally averaged rotation curve data, it can be seen more easily
if one separately  examines the northern (receding) and southern (approaching) sides of the galaxy \cite{bor,gen}.
For this reason, these and other authors, e.g. \cite{stacy} have argued that the wiggle is significant and 
demands an explanation.

For the purposes of the present  analysis, though, we shall for simplicity assume azimuthal symmetry.
In solving Eq.(\ref{9z}), we therefore set $\Sigma_{gas}$
to be the average of the northern and southern gas density measurements, $\Sigma_{gas}(r) = [\Sigma^{North}_{gas}(r)
+ \Sigma^{South}_{gas}(r)]/2$.
In figure 5b we compare the predicted rotation curve with the data. 
As the figure shows,
dissipative dark matter does a good job in reproducing both the rotation curve ($\chi^2_{red} = 0.45$) 
as well as the small wiggle at $r \approx 5$ kpc.
The phenomenological cored halo models also do a good job in fitting the rotation curve,
with $\chi^2_{red} = 0.30$ for quasi-isothermal and $\chi^2_{red} =  0.33$ for Burkert profile. 
[The NFW profile, on the other hand, gives a much poorer fit to the rotation curve data \cite{gen}.]
However such
smooth profiles cannot explain the apparent wiggle at around $r \approx 5$ kpc. 


We next consider 
a selection of dwarf galaxies from the LITTLE THINGS survey \cite{oh}.
That reference contains both the gas density and rotation curve data.
The results of the analysis are shown in figures 6-14 for the galaxies:
DDO87, DDO126, DDO133, DDO52, DDO47, DDO70,
NGC 3738, NGC 2366 and DDO154.
The figures show reasonable agreement with the data, especially as only one parameter, 
$\lambda'$, was adjusted.
The approach here also seems capable of explaining the subtle wiggles in the rotation curves and
could easily be improved by taking into account the azimuthal 
dependence of the gas density.
The galaxy NGC 1560, where the wiggle in
both the gas density and rotation curve at $r \approx 5$ kpc is much more pronounced in the northern half of the galaxy
\cite{bor,gen},
might be a good example to use if one were to take into account the azimuthal dependence.

The results in figures 5-14 assumed that the Kennicutt-Schmidt type relation had the exponent fixed at $N=2.0$.
Allowing $N$ to vary with values greater than $2.0$ generally gave better agreement between the predicted
rotation curves and the data. We illustrate this in figure 15, where we consider the predicted rotation curve
for DDO154 with varying $N$ values (where in each case we adjusted $\lambda'$ 
to give the correct normalization).
Given the uncertainties in the (large) star
formation rates, large $N$ values such as $N=3.0$ might be possible. However, we do not expect
Eq.(\ref{9z}) to be an exact description of the dark matter density 
profile.
In particular, the
assumptions of a) azimuthal symmetry b) isothermal halo and c) negligible dark stellar contribution
to the rotation curve, are all expected to be violated, at least to some extent, in specific cases. 
Given these potential uncertainties the level of agreement of the predicted rotation curves 
with the data is reasonable.

In addition to the galaxies shown, we also considered DDO101 and DDO168, which were
found to give much poorer fits and are not shown. 
DDO101 has a relatively low baryonic gas density and total baryonic mass. The resulting suppressed heating of
DDO101's halo may have been insufficient to support the halo from collapse.
That is, the dark halo of DDO101 is no longer a diffuse plasma, but may have collapsed into dark stars.
This is also supported by the relatively high measured value for the central density of its halo. 
By contrast,
DDO168 has a relatively high baryonic gas density in the central region. 
It is unclear why the formalism might fail for DDO168 but it might have something to do with this detail.
Also, the rotation curve data for both DDO168 (as with DDO70) show a turn over 
at the largest measured distances, which would require an understanding of the boundary region, beyond
the scope of this work.

\vskip 0.2cm
\centerline{\epsfig{file=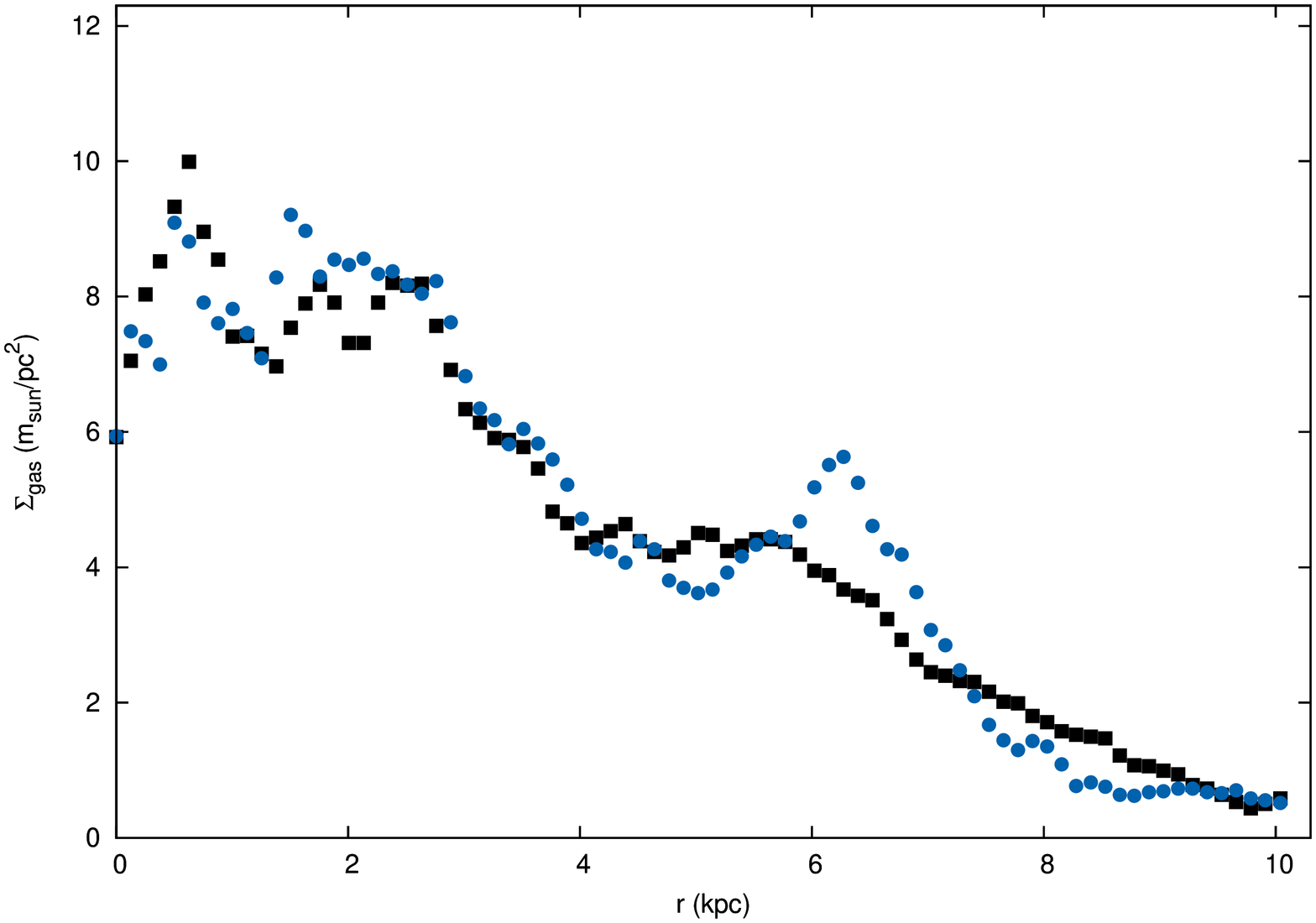,angle=270,width=11.7cm}}
\vskip 0.2cm
\noindent
{\small Fig. 5a: Baryonic gas density as found in \cite{gen}. The blue circles (black squares) represent an
average over the northern (southern) half of the galaxy.}

\vskip 0.2cm
\centerline{\epsfig{file=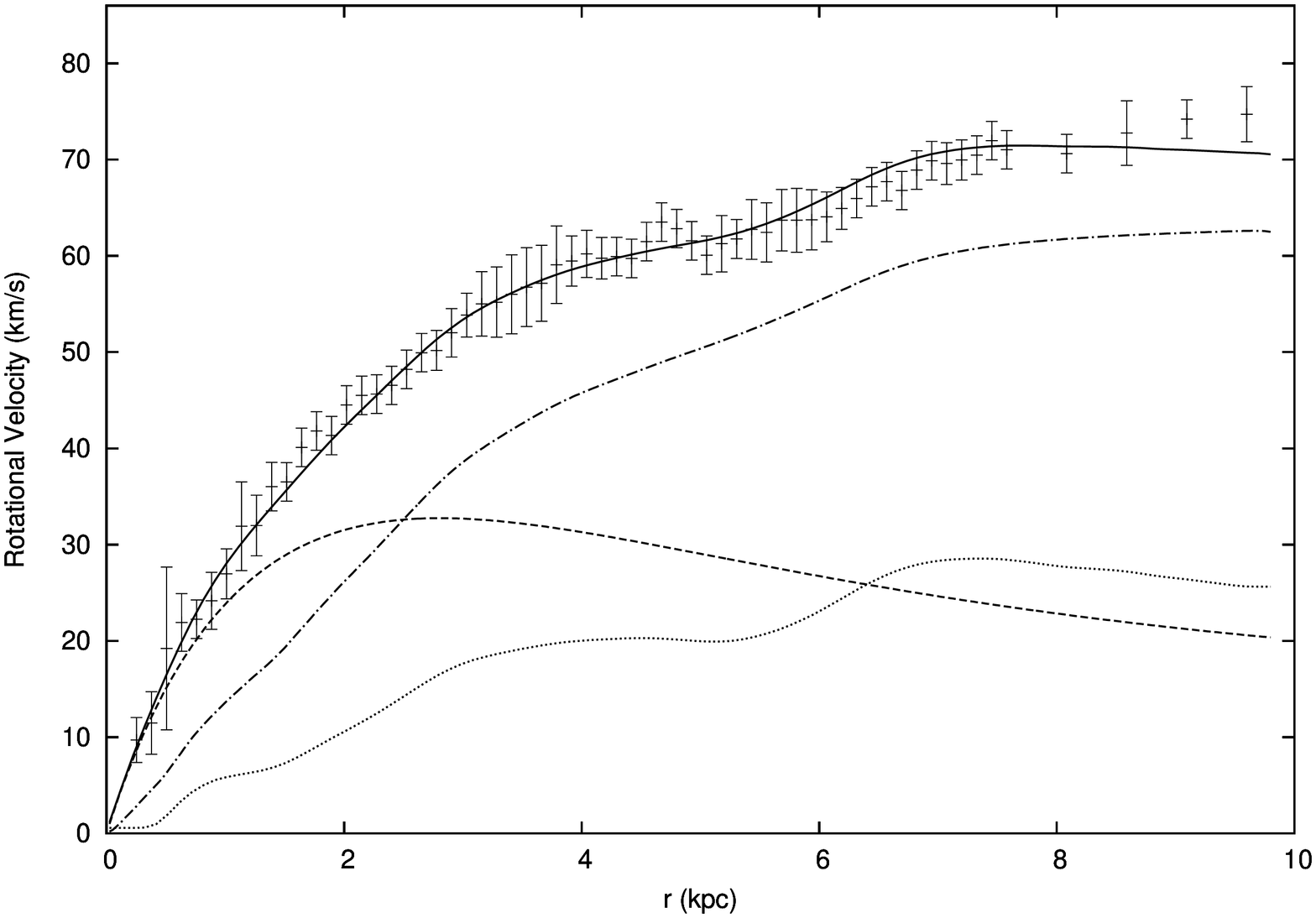,angle=270,width=11.7cm}}
\vskip 0.2cm
\noindent
{\small Fig. 5b: Predicted rotation curve for NGC 1560 [solid line].
Also shown are the separate contributions from
the assumed dissipative dynamics, Eq.(\ref{9z}) [dashed-dotted line], the baryonic stellar [dashed line] and baryonic gas
contributions [dotted line]. The data is from \cite{gen}.}

\vskip 0.8cm

\centerline{\epsfig{file=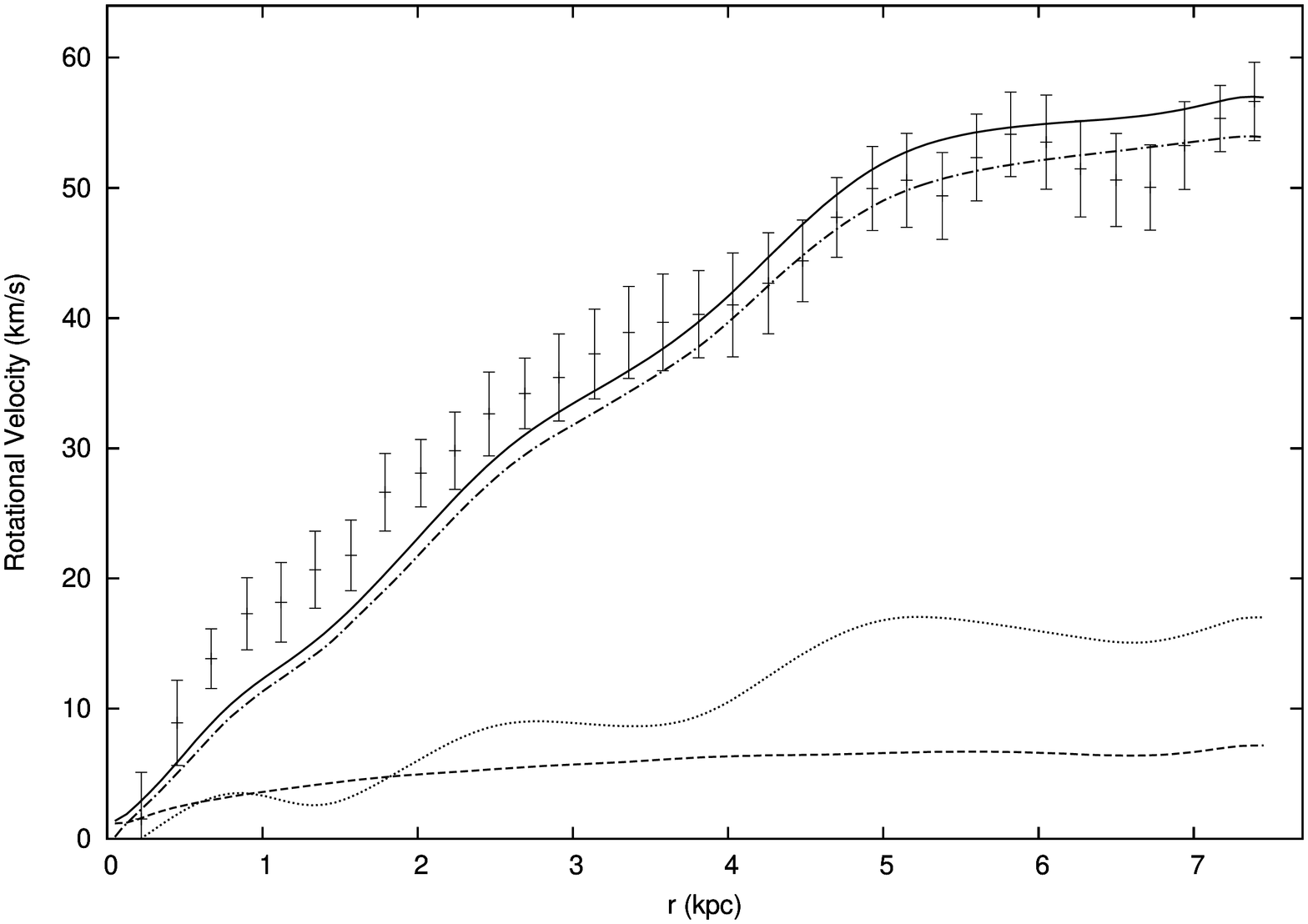,angle=270,width=11.7cm}}
\vskip 0.3cm
\noindent
{\small Fig. 6: Predicted rotation curve for DDO87 [solid line]. 
Also shown are the separate contributions from
the assumed dissipative dynamics, Eq.(\ref{9z}) [dashed-dotted line], the baryonic stellar [dashed line] and baryonic gas
contributions [dotted line]. The data is from \cite{oh}.}

\vskip 0.8cm
\centerline{\epsfig{file=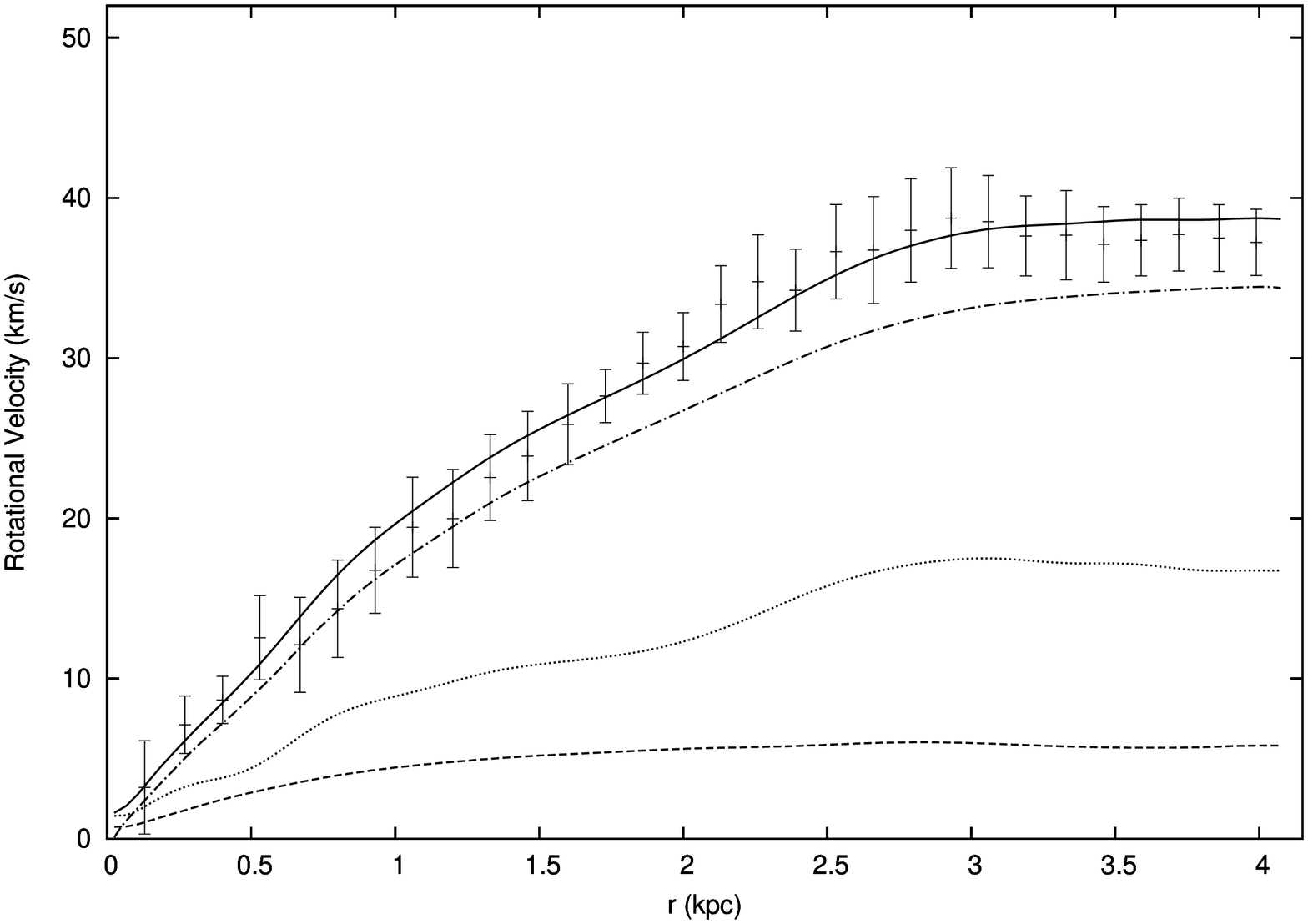,angle=270,width=11.7cm}}
\vskip 0.3cm
\noindent
{\small Fig. 7: Predicted rotation curve for DDO126 [solid line]. Notation as in Fig. 6, data from \cite{oh}.}

\vskip 0.8cm
\centerline{\epsfig{file=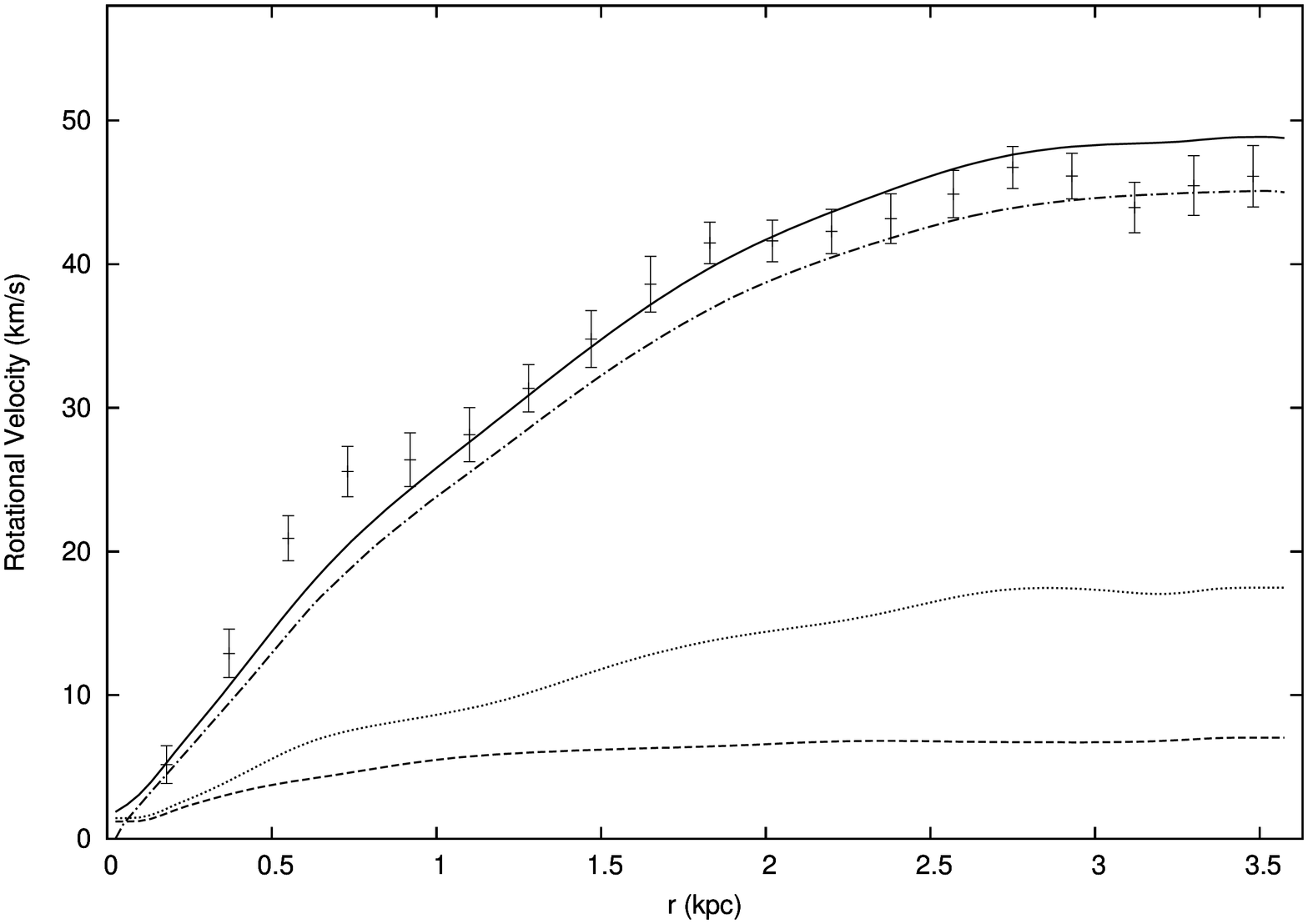,angle=270,width=11.7cm}}
\vskip 0.3cm
\noindent
{\small Fig. 8: Predicted rotation curve for DDO133 [solid line]. Notation as in Fig. 6, data from \cite{oh}.}

\vskip 0.8cm
\noindent
\vskip 0.2cm
\centerline{\epsfig{file=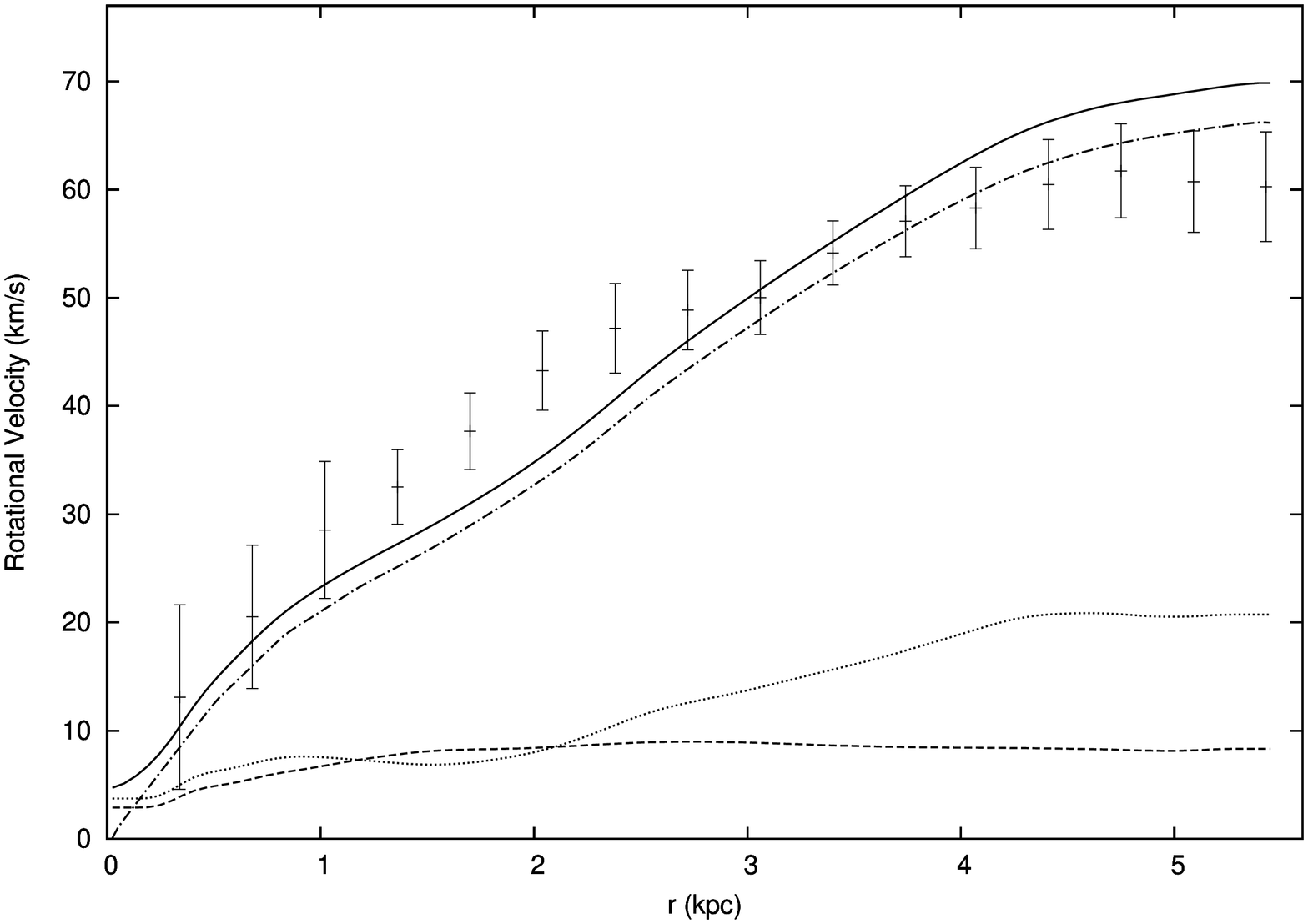,angle=270,width=11.7cm}}
\vskip 0.3cm
\noindent
{\small Fig. 9: Predicted rotation curve for DDO52 [solid line]. Notation as in Fig. 6, data from \cite{oh}.}
\noindent

\vskip 0.8cm
\centerline{\epsfig{file=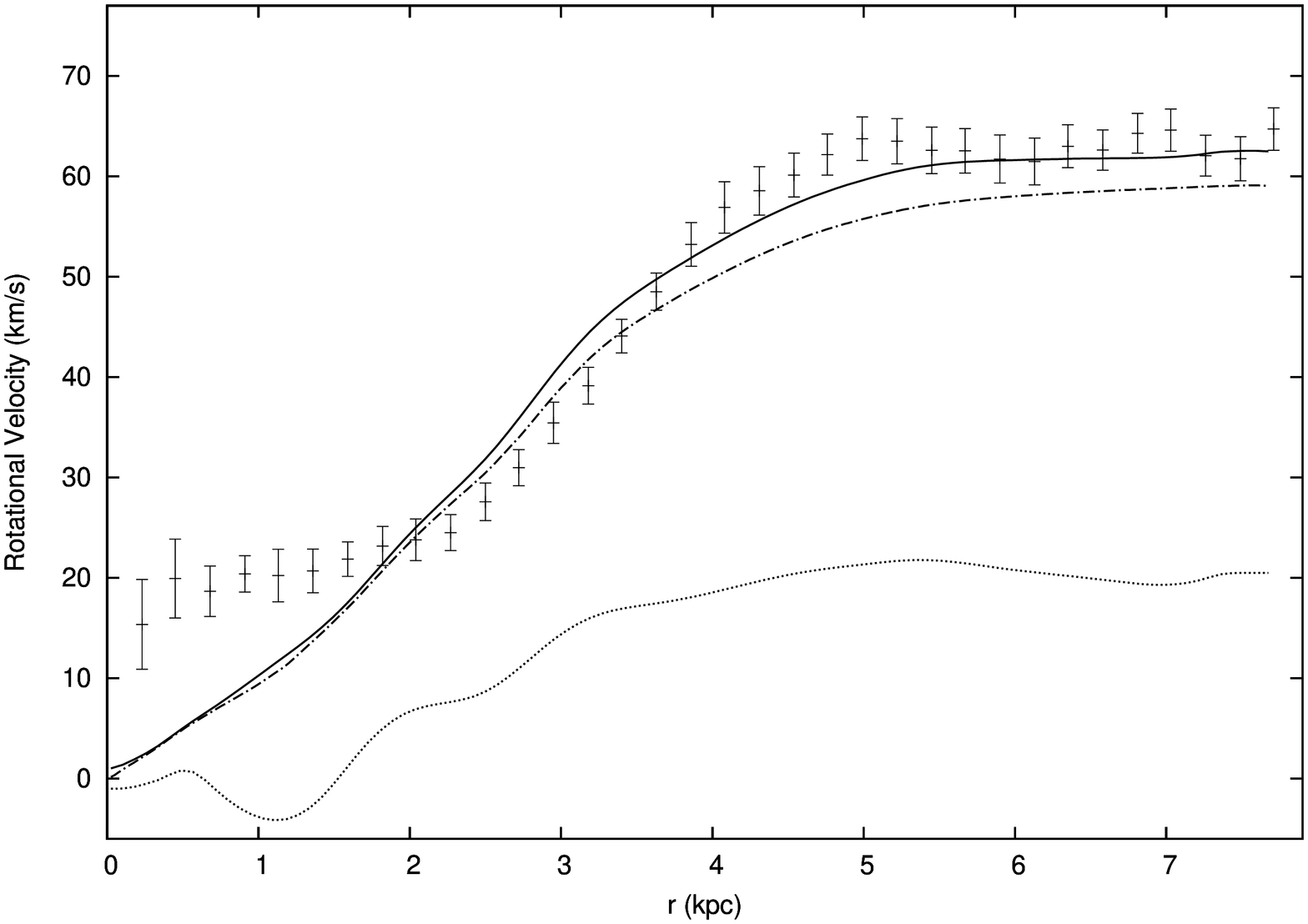,angle=270,width=11.7cm}}
\vskip 0.3cm
\noindent
{\small Fig. 10: Predicted rotation curve for DDO47 [solid line]. Notation as in Fig. 6, data from \cite{oh}.}

\vskip 0.8cm
\centerline{\epsfig{file=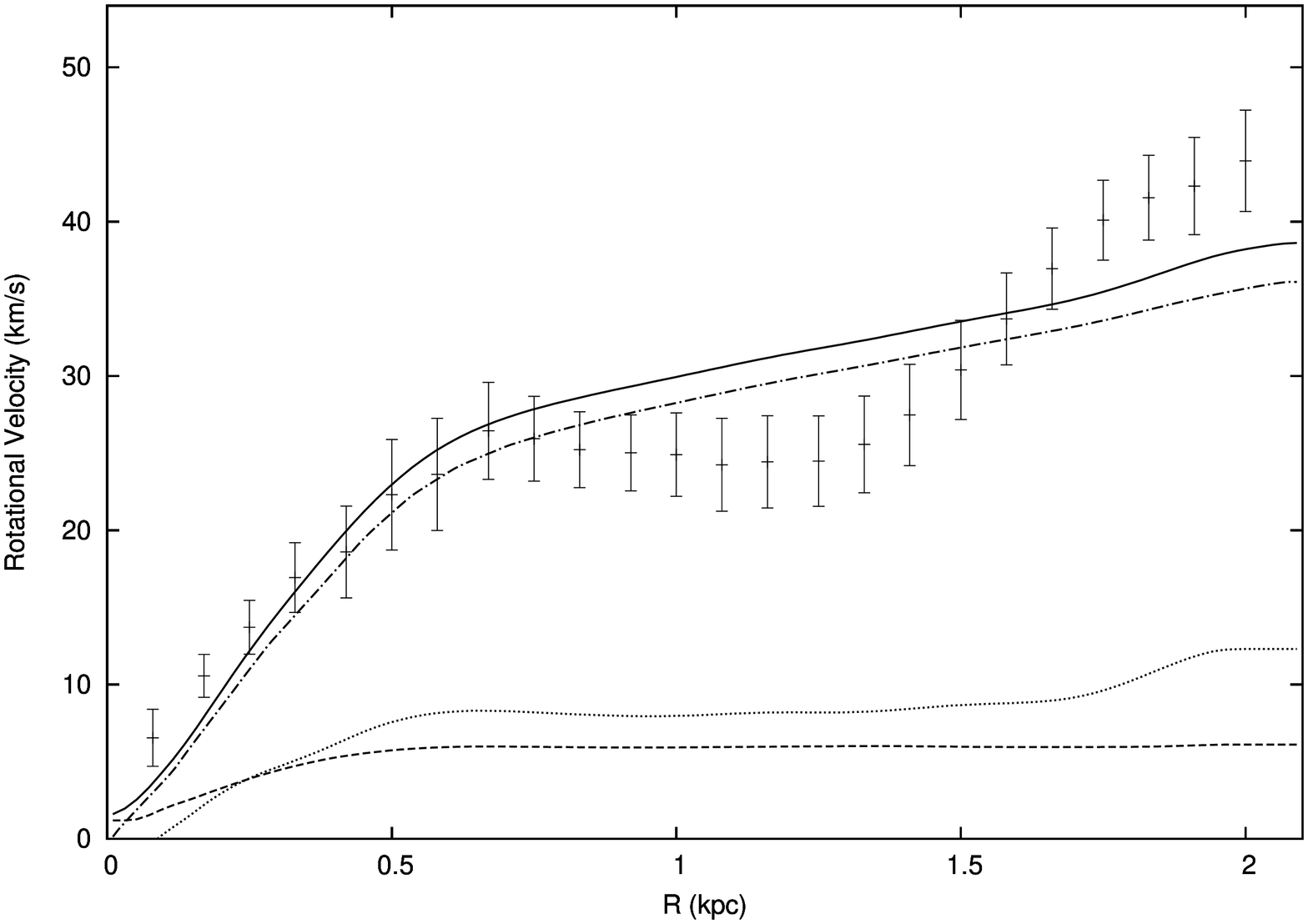,angle=270,width=11.7cm}}
\vskip 0.3cm
\noindent
{\small Fig. 11: Predicted rotation curve for DDO70 [solid line]. Notation as in Fig. 6, data from \cite{oh}.}

\vskip 0.8cm
\centerline{\epsfig{file=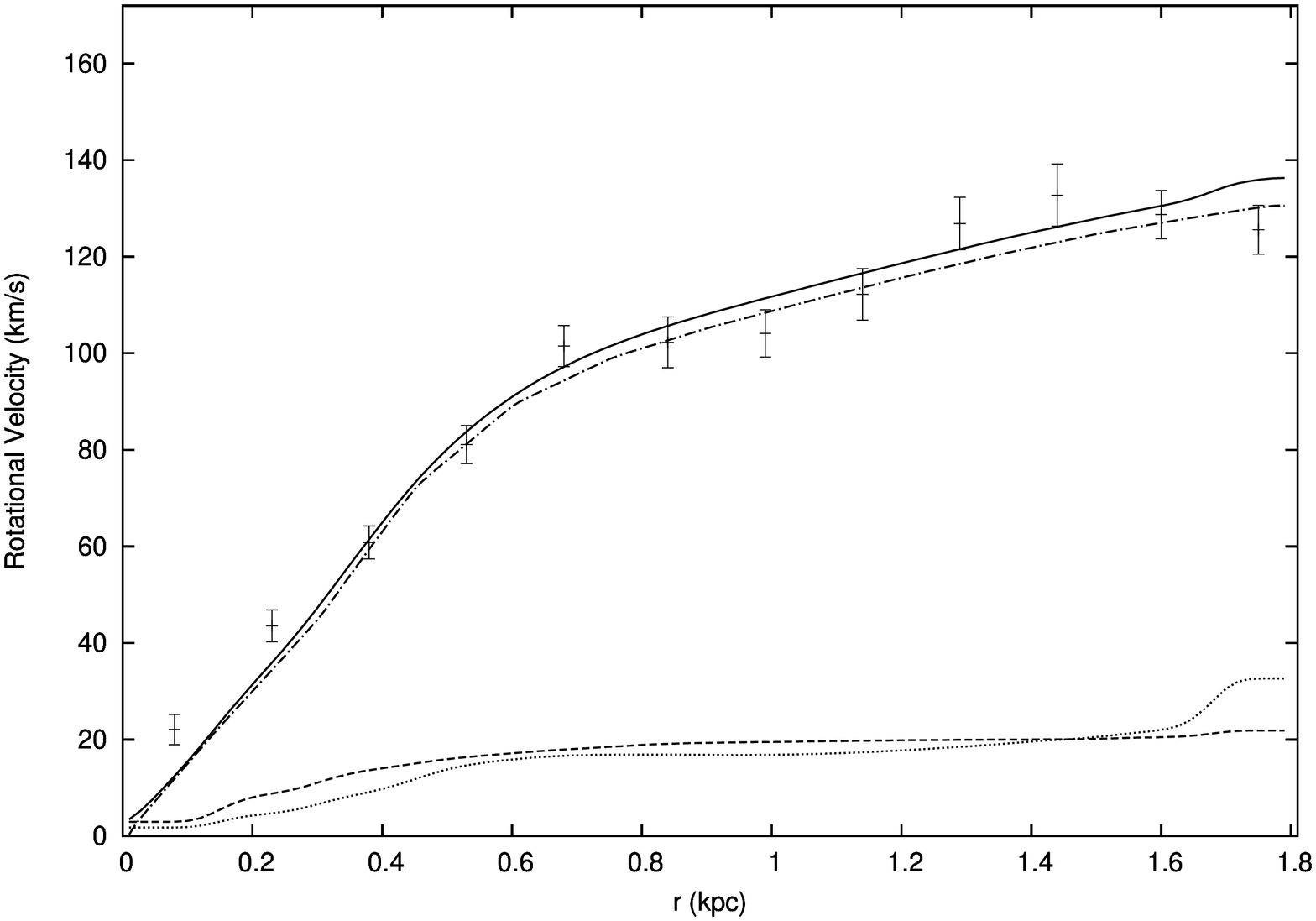,angle=270,width=11.7cm}}
\vskip 0.3cm
\noindent
{\small Fig. 12: Predicted rotation curve for NGC 3738 [solid line]. Notation as in Fig. 6, data from \cite{oh}.}
\vskip 0.8cm
\centerline{\epsfig{file=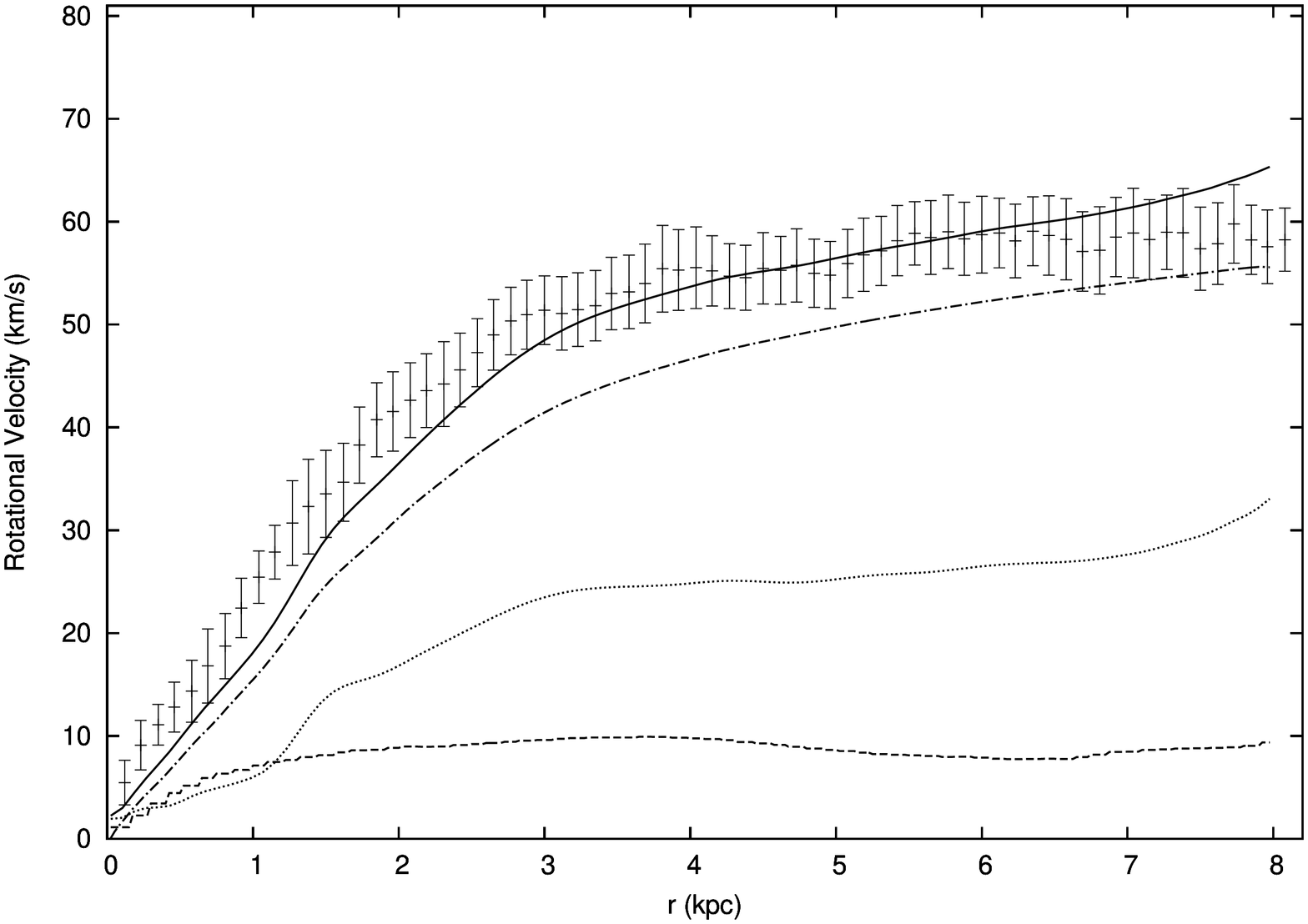,angle=270,width=11.7cm}}
\vskip 0.3cm
\noindent
{\small Fig. 13: Predicted rotation curve for NGC 2366 [solid line]. Notation as in Fig. 6, data from \cite{oh}.}

\vskip 0.8cm
\centerline{\epsfig{file=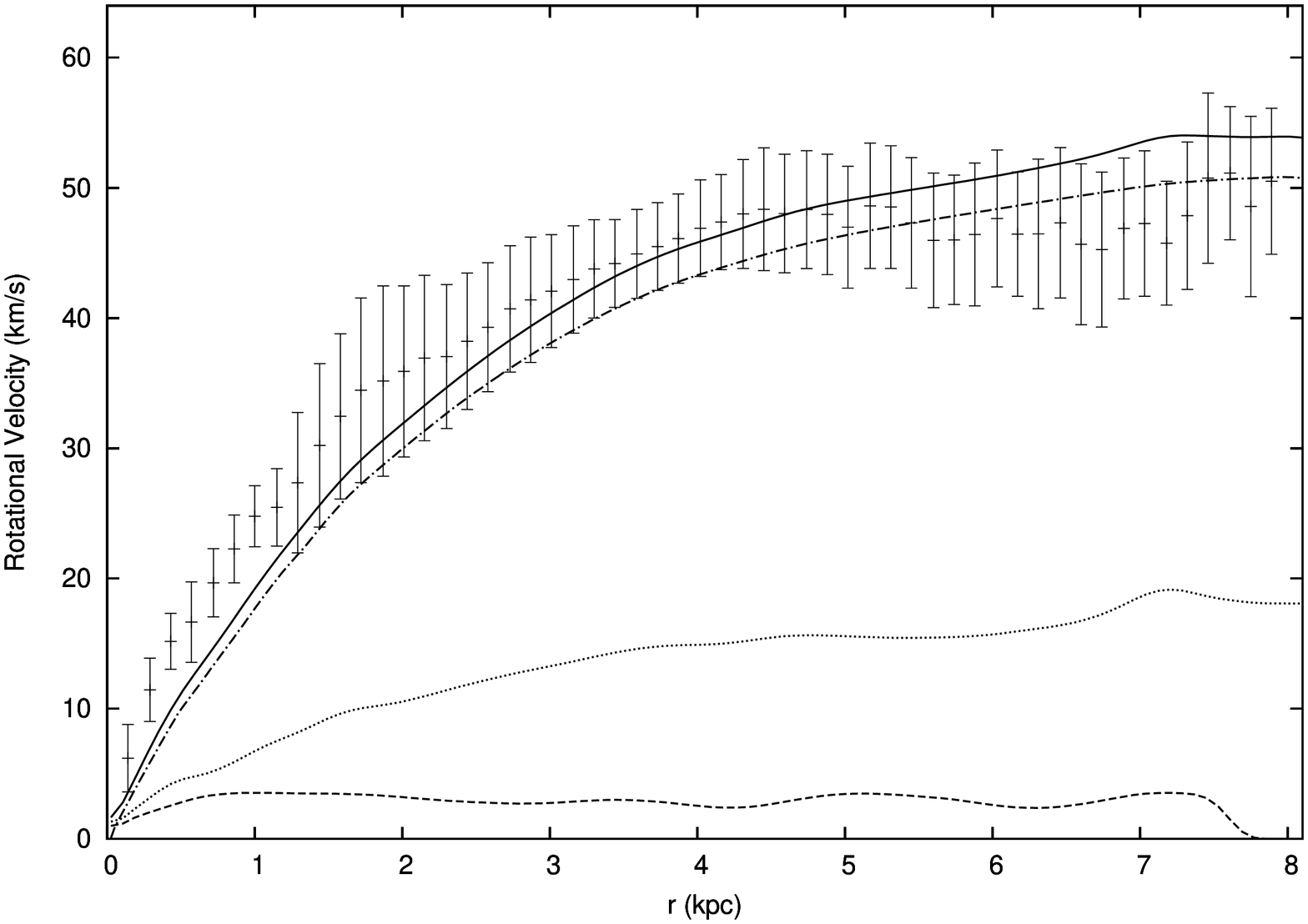,angle=270,width=11.7cm}}
\vskip 0.3cm
\noindent
{\small Fig. 14: Predicted rotation curve for DDO154 [solid line]. Notation as in Fig. 6, data from \cite{oh}.}

\vskip 0.8cm
\centerline{\epsfig{file=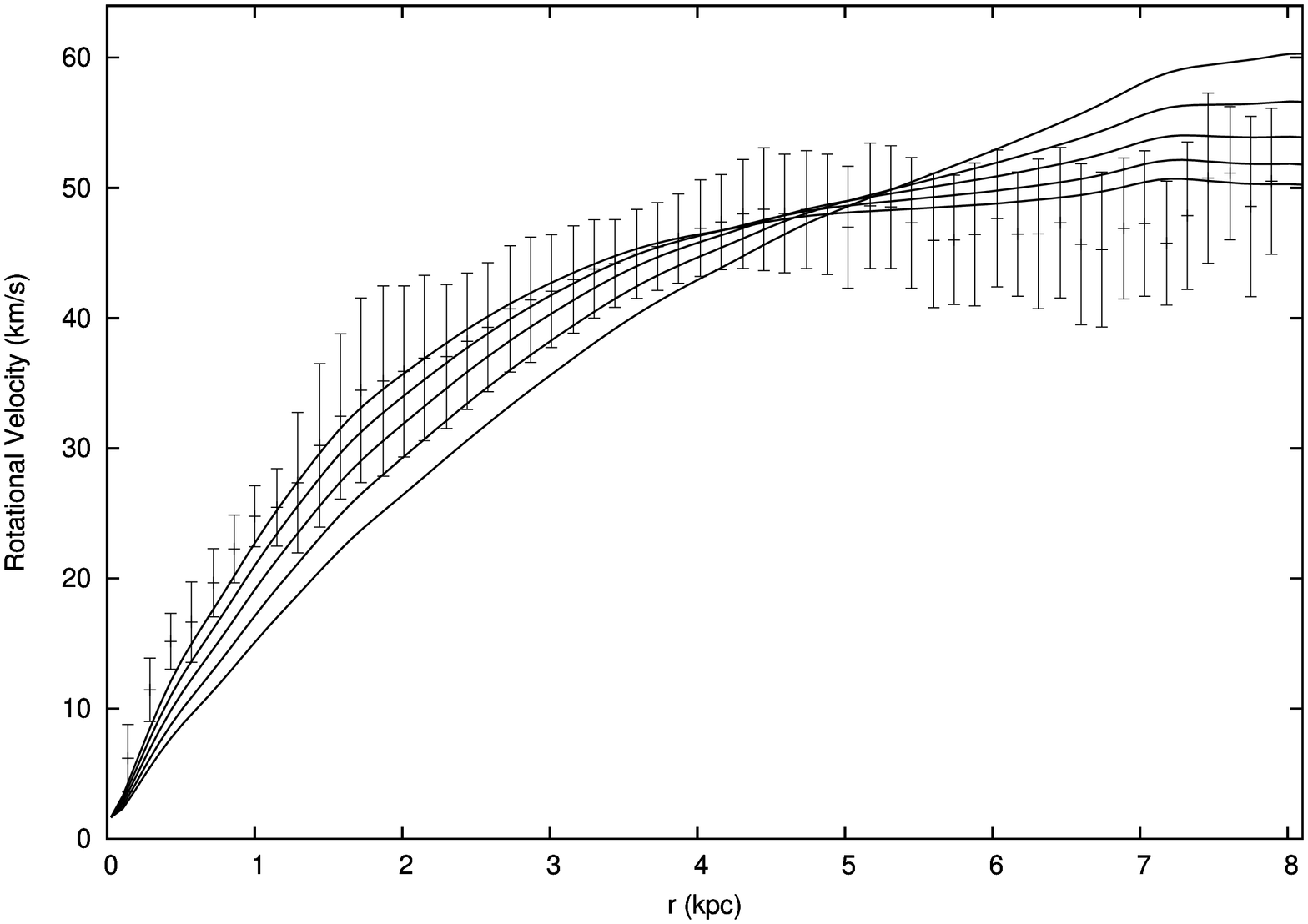,angle=270,width=11.7cm}}
\vskip 0.3cm
\noindent
{\small Fig. 15: Predicted rotation curve (dark matter + baryons) for DDO154, with Kennicutt-Schmidt exponent, $N$, varying.
Curves
from (left) bottom to top are for $N=1.0, \ N=1.5, \ N=2.0, \ N=2.5, \ N=3.0$.
}
\vskip 0.8cm


\section{Conclusions}

We have considered dissipative dark matter, in which dark matter particles forming galactic halos are presumed to 
have dissipative self-interactions.
Hidden sector gauge models featuring dissipative dark matter are straightforward to construct. An
example being hidden sector dark matter charged under an unbroken $U(1)'$ gauge symmetry, leading to dark matter
interacting with a massless gauge boson - the dark photon. The dark photon not only mediates
self interactions among the dark sector particles, but allows for halo cooling via dissipative 
particle processes such as dark bremsstrahlung. 
Mirror dark matter, where the hidden sector is an exact duplicate of the standard model sector, is
a constrained example of such a theory.
Many other dissipative dark matter
models may be possible, e.g. where the dark photons are light rather than massless, or even where they
are replaced by light scalar particle(s).

If dark matter is dissipative then the dark matter halo around disk galaxies can have quite nontrivial behavior.
In addition to cooling, the halo can have heating powered by ordinary type II supernovae in the disk.
Naturally, this requires a small coupling between the dark matter particles and ordinary matter ones,
with the photon - dark photon kinetic mixing interaction a likely suspect. 
With such an interaction ordinary supernovae become a major source of dark photons which can supply
a substantial amount of energy to the halo.

This type of dark matter, which can be modeled as a fluid satisfying Euler's equations of fluid dynamics,
will expand and contract in response to these heating and cooling influences.
In the current epoch, the halo is anticipated to have evolved into a steady-state configuration where the energy radiated
at each point approximately balances the heating. If this indeed occurs, then the physical properties of the halo
will be strongly constrained and
a very simple formula, Eq.(\ref{3z}), relating the dark matter density with the supernova
energy sources results.

For  a given supernova distribution, a prediction for the rotation curves of individual galaxies follows - all of the 
model dependence, kinetic mixing parameter, relevant cross-sections etc can be condensed into a single parameter.
The spatial distribution of supernovae was modelled with a Freeman disk profile and also
via a Kennicutt-Schmidt-type power law. 
The resulting rotation curves were then compared with observations and found to be in reasonably good agreement.
In particular, subtle `wiggles' in the rotation curve which appear to be correlated with corresponding features
in the disk gas distribution could plausibly be explained in this approach.


\vskip 1cm
\noindent
{\large Acknowledgments}

\vskip 0.2cm
\noindent
The author would like to thank S. Vagnozzi for comments on a draft of this paper and also
thank: S. Oh, G. Gentile and W. Blok
for making available rotation curve data files.
This work was supported by the Australian Research Council.

\end{document}